\documentclass[pra, twocolumn,amsmath, amssymb, notitlepage, longbibliography, showpacs]{revtex4-1}

\usepackage{dsfont}
\usepackage[english]{babel}
\usepackage{graphicx}
\usepackage[usenames, dvipsnames]{color}
\usepackage{rotating}
\usepackage[breaklinks=true]{hyperref}
\hypersetup{
  colorlinks   = true, 
  urlcolor     = blue, 
  linkcolor    = blue, 
  citecolor   =  red 
}
\usepackage{soul}
\usepackage{xr}
\makeatletter
\def\@bibdataout@aps{%
 \immediate\write\@bibdataout{%
  @CONTROL{%
   apsrev41Control,author="08",editor="1",pages="0",title="0",year="1",eprint="1"%
  }%
 }%
 \if@filesw
  \immediate\write\@auxout{\string\citation{apsrev41Control}}%
 \fi
}%
\makeatother 

\graphicspath{ {./Pictures/} }

\newcommand{\bra}[1]{\ensuremath{\left\langle #1\r|}}
\newcommand{\ket}[1]{\ensuremath{\left|#1\r\rangle}}

\newcommand{\mean}[1]{\ensuremath{\left\langle #1\r\rangle}}

\newcommand{\cc}{^{\ast}}							
\newcommand{\hc}{^{\dagger}}							

\newcommand{\ee}{\mathrm{e}}						
\newcommand{\ii}{\mathrm{i}}			             			

\newcommand{\comm}[2]{\left[ #1, #2 \right]} 				
\newcommand{\nn}{\nonumber}							
\newcommand{\abss}[1]{\ensuremath{ \left| #1 \right|^{2} }}	
\newcommand{\diss}[1]{\mathcal{D}[ #1 ]}					
\renewcommand{\l}[0]{\left}
\renewcommand{\r}[0]{\right}

\newcommand{\Tr}{\text{Tr}}

\newcommand{\eq}[1]{\eqref{#1}}

\newcommand{\eqn}[1]{Eq.~\eq{#1}}
\newcommand{\eqns}[1]{Eqs.~\eq{#1}}
\newcommand{\fig}[1]{Fig.~\ref{#1}}

\begin{document}

\title{Nonreciprocal  Atomic Scatterering: A saturable, quantum Yagi-Uda antenna }

\author{Clemens M\"uller}
\email{c.muller2@uq.edu.au}
\affiliation{ARC Centre of Excellence for Engineered Quantum Systems, School of Mathematics and Physics, The University of Queensland, Saint Lucia, Queensland 4072, Australia}

\author{Joshua Combes}
\affiliation{ARC Centre of Excellence for Engineered Quantum Systems, School of Mathematics and Physics, The University of Queensland, Saint Lucia, Queensland 4072, Australia}

\author{Andr\'es Rosario Hamann}
\affiliation{ARC Centre of Excellence for Engineered Quantum Systems, School of Mathematics and Physics, The University of Queensland, Saint Lucia, Queensland 4072, Australia}

\author{Arkady Fedorov}
\affiliation{ARC Centre of Excellence for Engineered Quantum Systems, School of Mathematics and Physics, The University of Queensland, Saint Lucia, Queensland 4072, Australia}

\author{Thomas M. Stace}
\email{stace@physics.uq.edu.au}
\affiliation{ARC Centre of Excellence for Engineered Quantum Systems, School of Mathematics and Physics, The University of Queensland, Saint Lucia, Queensland 4072, Australia}


\date{\today}

\begin{abstract}
	Recent theoretical studies of a pair of atoms in a 1D waveguide find that the system responds asymmetrically to incident fields from opposing directions at low powers.  
	Since there is no explicit time-reversal symmetry breaking elements in the device, this has caused some debate.  
	Here we show that the asymmetry arises from the formation of a quasi-dark-state of the two atoms, which saturates at extremely low power.  
	In this case the nonlinear saturability explicitly breaks the assumptions of the Lorentz reciprocity theorem.
	Moreover, we show that the  statistics of the output field from the driven system can be explained by a very simple stochastic mirror model and that at steady state, 
	the two atoms and the local field are driven to an entangled, tripartite $\ket{W}$ state. 
	Because of this, we argue that the device is better understood as a saturable Yagi-Uda antenna, a distributed system of differentially-tuned dipoles that couples asymmetrically to external fields. 
\end{abstract} 



\maketitle


Nonreciprocal devices, such as isolators, circulators, and gyrators, are important components for optical and microwave  technologies. 
They are typically used to route or isolate signals propagating in different directions. 
Recently, a unidirectional, two-atom device has been identified as potentially useful in quantum electronics \cite{FratMascSafa14, DaiRoulLe15, MascSantAuff16, FratGhob16, GonzMoreGarc16, OrdoEzzaJoul17,FangBara17}, 
building on earlier analyses of distributed atomic systems~\cite{ RudoFiceDalt95, RudoFice98, MakaLeto03, LembRsheAldo13, LaakPlet14}.  
Transmission through this device depends asymmetrically on the direction of the input field, hence it has been dubbed a \emph{quantum diode}.

\begin{figure}[t]
	\centering
	\includegraphics[width=\columnwidth]{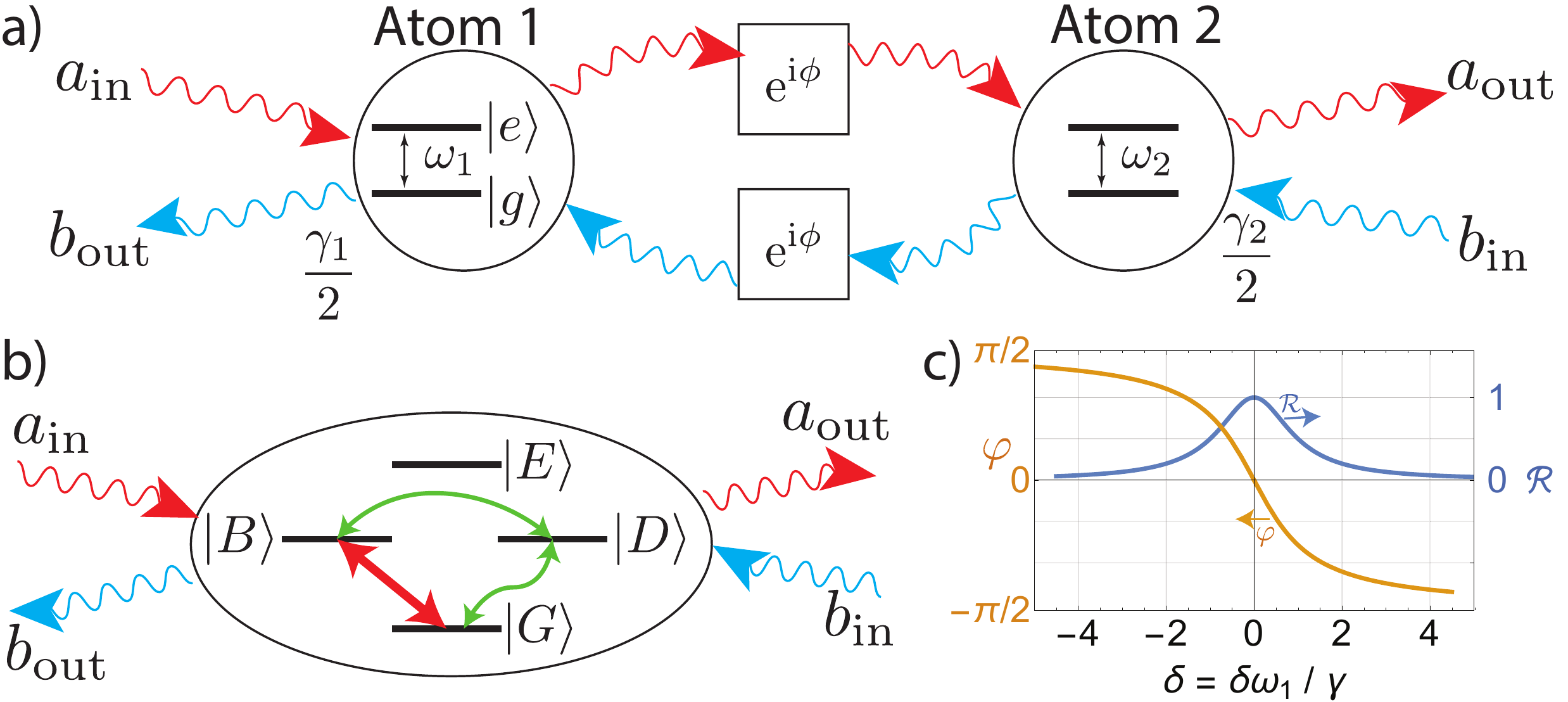}
	\caption{(colour online) (a) Schematic illustration of two atoms in a wave-guide, driven by a field incident from the left with amplitude $\mean{a_{\rm in}} = \alpha$, or from the right with \mbox{$\mean{b_{\rm in}}=\beta$}. 
		(b) Hybridised model, in which an incident field  scatters strongly (i.e.\ reflects) off the $\ket{G}=\ket{gg}\leftrightarrow\ket{B}\sim\ket{eg}-\ket{ge}$ transition.  
		For a field incident from the left, interference between direct and indirect excitation channels couple weakly to the dark state $\ket{D}\sim\ket{eg}+\ket{ge}$.  
		Driving from the right does not couple to $\ket{D}$. (c) Reflectance $\mathcal R$ and phase shift $\varphi$ versus dimensionless detuning, $\delta=\delta\omega_1/\gamma$, for a field reflected off a single atom.
	}
	\label{fig:setupschematic}
\end{figure}

The quantum diode consists of a pair of spatially-separated, nondegenerate atoms  in a 1D waveguide,  shown in \fig{fig:setupschematic}a,  tuned to  discriminate between a coherent field $\alpha$ incident from the left, and a coherent field $\beta$ incident from the right.  
\emph{Prima facie}, this appears to violate reciprocity: the transmission coefficients of a passive, linear, time-reversal-symmetric scatterer should satisfy $\mathcal T_\leftarrow = \mathcal T_\rightarrow$, so there is an interesting  question as to the origin of the transmission asymmetry.

Here, we derive a master equation for the  driven two-atom system shown in \fig{fig:setupschematic}a. 
We show that the two-atom dark state~\cite{FangBara17} responsible for the asymmetry arises from entanglement between the matter and the field~\cite{MakaLeto03}. 
This leads to non-reciprocal~\cite{JalaPetrEich13} and incoherent~\cite{FangBara17} scattering matrices, and we establish the maximum possible `diode efficiency'~\cite{DaiRoulLe15} of $2/3$, for which the steady state is inverted. 
Finally, we show that a toy-model of a randomly fluctuating mirror replicates the  statistics of the scattered field and corresponds exactly to the rate equation model when adiabatically eliminating all coherences.  

The picture that emerges is that in the steady-state, under cw-driving from one direction, the two atoms become entangled with the local electromagnetic field  in a tripartite $\ket{W}$ state.  
In the atomic Hilbert space, this corresponds to a long-lived, probabilistic mixture of the ground and  dark  states. Since scattering arises from coherence between the ground and bright states, the dark state population effectively decouples the scatterer from the  field, resulting in non-zero transmission. 
In contrast, driving from the opposite direction does not couple to the dark state at all. Based on these observations, we argue that the device should be understood as a saturable  Yagi-Uda antenna (a directional dipole array)~\cite{Kosako:2010aa,LembRsheAldo13}, 
and we speculate that non-reciprocity may be enhanced in an $n>2$ atom device.  

This paper is organised as follows: We introduce the master equation describing the two atoms and the coherent drive via the waveguides in Section~\ref{sec:System} and then start the analysis in Section~\ref{sec:SS} by calculating the steady-state of the two atoms under driving in the parameter regime relevant for rectification. 
These results then motivate the division of the two atom Hilbert space into a fast and slow subspace and we derive the dynamical equations for the scattering when adiabatically eliminating the fast subspace in the following Section~\ref{sec:Adiab} and discuss the scattering characteristics of this system in Section~\ref{sec:Scatter}. 
Another step of adiabatically eliminating the remaining coherences in the slow subspace then leads to a toy model of a ``flapping'' mirror, which we explore in section~\ref{sec:FM} before discussing the rectification properties of this device in Section~\ref{sec:Rect}.
The appendix containfes details of the calculations and derivations.

\section{System \label{sec:System}}
We model the system of two two-level atoms depicted in \fig{fig:setupschematic}b, bi-directionally cascaded in a 1D waveguide~\cite{Zeeb:PRA:2015, CombKercSaro16} (also see appendix~\ref{app:slh}).  
The atoms are driven by a coherent field at frequency $\omega_c$, and separated by a distance $d$ (with corresponding phase shift $\phi=\omega_c d/c_s=2\pi d/\lambda_c$ \cite{LaluSandLoo13}).  
The device operates near the first resonance, for which the inter-atomic spacing is  half a wavelength, i.e.\ $\phi\approx\pi$. The  evolution of the system in the local atomic basis $\{{\ket{gg}},{\ket{ge}},{\ket{eg}},{\ket{ee}}\}$ is described by Hamiltonian terms and dissipators 
\begin{align}
	H_k =  - \omega_k\, \sigma_z^{(k)} /2\,,\quad   
	L_k = \sqrt{{\gamma_k}/2}\, \sigma_-^{(k)} \,, \label{eqn:lk}
\end{align}
with $k=1,2$ for the two atoms in the waveguide, and $ \sigma_-={\ket{g}}{\bra{e}}$.  
The master equation for the two-atom density matrix is
\begin{align}
	\dot{\rho} =\mathcal{L}\rho\equiv  -\ii \comm{H_T}{\rho} + \diss{L_\rightarrow}\rho + \diss{L_\leftarrow}\rho,
	\label{eq:ME0}
\end{align}
where $\diss{X}\rho=X\rho X\hc-\tfrac12(X\hc X\rho +\rho X\hc X)$, and 
\begin{align}
	H_T =& H_1 + H_2 -\ii 
	\big( \alpha L_1\hc - \alpha\cc L_1 \big)/2 -\ii 
	 \big( \beta L_2\hc - \beta\cc L_2 \big)/2 \nn\\
		&-\ii 
		 \big( \ee^{\ii \phi} L_2\hc (L_1 + \alpha) - \ee^{-\ii \phi} (L_1\hc + \alpha\cc) L_2 \big)/{2} \nn\\
		&-\ii 
		\big( \ee^{\ii \phi} L_1\hc (L_2 + \beta) - \ee^{-\ii \phi} (L_2\hc + \beta\cc) L_1 \big)/{2} \,, \label{eq:HSLH}\\
	L_\rightarrow =& L_2 + \ee^{\ii \phi} L_1 + \ee^{\ii \phi} \alpha \,, \label{eq:L1SLH}\\
	L_\leftarrow =& L_1 + \ee^{\ii \phi} L_2 + \ee^{\ii \phi} \beta \,. \label{eq:L2SLH}
\end{align}
Terms like $L_2\hc L_1$ in $H_T$ represent effective inter-atomic coupling, induced by their mutual coupling to the waveguide.

The left-moving output field amplitude and photon flux are, respectively
\begin{align}
	\alpha_{\text{out}}(t)&= \langle a_{\text{out}}(t) \rangle =  \text{Tr}\l\{  L_\leftarrow \rho \r(t)\} , \nn\\
	\mathcal{A}_{\text{out}}(t) &= 	\langle a_{\text{out}}^\dagger a_{\text{out}}(t) \rangle =  \text{Tr}\l\{  L\hc_\leftarrow L_\leftarrow \rho(t) \r\}.
\end{align}
The right-moving amplitude, $\beta_{\text{out}}$, and flux, $\mathcal{B}_{\text{out}}$, depend similarly on $L_\rightarrow $. 
Without loss of generality, we consider the two atoms to be coupled symmetrically to the waveguide, so that $\gamma_{1} = \gamma_{2} = \gamma$.

Atom 2 (depicted in \fig{fig:setupschematic}) is resonant with the carrier frequency, $\omega_c$, and to break inversion symmetry, we detune atom 1 by an amount \mbox{$\omega_1-\omega_c=\delta\omega_1\equiv-\delta \,\gamma$}.  
In linear response, the reflectance of each atom is a Lorentzian in the dimensionless detuning, $\delta$, as shown in \fig{fig:setupschematic}c, and there is a corresponding  phase shift, $\varphi$, in transmission.  For small detuning, this phase shift is $\varphi\approx-\delta$.  
We perturb the geometric inter-atomic separation to  compensate for this phase shift, so that $\phi=\pi-\delta$.  
This choice of $\delta\omega_1$ and $\phi$ is consistent with Ref.~\cite{DaiRoulLe15} and, as shown in appendix~\ref{app:optimum}, optimises the asymmetry in the response of the system. 
In what follows, we adopt units where $\gamma=1$.

\section{Steady state \label{sec:SS}}
As we are interested in the scattering properties of the two-atom system, we start our analysis by calculating the steady-state of the atoms under driving from either left or right. 
The results of this section will guide the subsequent analysis and allow us to identify a reduced slow subspace relevant for the scattering dynamics, and which will enable us to  adiabatically eliminate fast degrees of freedom from the cascaded master equation in Section~\ref{sec:Adiab}. 

We solve for the steady-state of the master equation, $\mathcal L\rho_{ss}^{(\alpha,\beta,\delta)} =0$, perturbatively  in $\delta$, using the expansions
\begin{eqnarray}
\rho_{ss}^{(\alpha,\beta,\delta)} &=& \bar\rho_{0} + \ii \delta \bar \rho_{1} + \delta^{2} \bar \rho_{2} +\ldots,\nonumber\\
\mathcal L &=& \mathcal L_{0} + \ii \delta \mathcal L_{1} + \delta^{2} \mathcal L_{2} + \ldots.
\end{eqnarray}
This expansion assumes that $\delta$ is the smallest quantity in the problem, consistent with earlier treatment of this problem~\cite{DaiRoulLe15, FangBara17}. 
Further, we assume that driving is far below the saturation power for each individual atom, so that \mbox{$\delta\ll|\alpha|+|\beta|\ll \gamma=1$}.  This allows us to make analytic progress, and, as we show is the regime in which interesting physics occurs.

The solution to the zeroth-order equation, $0 = \mathcal L_{0} \bar\rho_{0}$ is the nullspace of the superoperator $\mathcal L_0$,
which is two-fold degenerate,
\begin{align}
	\bar \rho_{0} = p_G(\alpha,\beta,\delta)\, \bar \rho_{0}^{(1)} + p_D(\alpha,\beta,\delta)\, \bar \rho_{0}^{(2)} \,, \label{eqn:ss0}
\end{align}
where $\bar \rho_{0}^{(1)}={\ket{G}}{\bra{G}}+O(\alpha^2,\beta^2)$ and $\bar \rho_{0}^{(2)}={\ket{D}}{\bra{D}}$,  $\ket{G}={\ket{gg}}$,  \mbox{$\ket{D}=(\ket{ge}+\ket{eg})/\sqrt{2}$} is the `dark' state, and  $p_{G/D}$ are as-yet undetermined  coefficients.   
As shown in appendix~\ref{app:symmbasis}, the system thus hybridises into the symmetric ($\ket{D}$) and antisymmetric ($\ket{B}$) states shown in \fig{fig:setupschematic}b, in which the steady state is well-approximated by a probabilisitic mixture of the ground state and the dark state. 

We calculate the higher-order corrections, $\bar \rho_{j}$, by a generalised nullspace analysis of the higher-order expansions of $\mathcal L \bar\rho =0$ (see appendix~\ref{app:ss}). At second order we find that $p_G$ and $p_D$ are related by
\begin{align}
	p_{D}/p_G=\left\{\begin{array}{lc}  2+ \alpha^{2} + 2\alpha^{4}  & \textrm{for $\alpha$ driving ($\beta_{\textrm{in}}=0$)} \\
	 \beta^2+2\beta^4 & \textrm{for $\beta$ driving ($\alpha_{\textrm{in}}=0$)}  \end{array}\right..
	\label{eq:pDG}
\end{align}
Together with normalisation, \mbox{$\Tr\{\bar \rho_{0}\}=p_G+p_D=1$}, we find
\begin{align}
	\rho_{ss}^{(\alpha,0,\delta)}&=\tfrac{1}{3}{\ket{G}}{\bra{G}}+\tfrac23{\ket{D}}{\bra{D}}+O(\alpha^2),\nonumber\\
	\rho_{ss}^{(0,\beta,\delta)}&={\ket{G}}{\bra{G}}+O(\beta^2).\label{eqn:rhoss} 
\end{align}
Thus, for $\alpha$-driving (i.e.\ from the left), the steady-state of the system is dominated by the dark state, whereas  $\beta$ driving (i.e.\ from the right) is decoupled from the dark state, and leaves the atoms in the ground state.  These results are apparently independent of the driving amplitudes.  For $\alpha$ driving, this arises because the dark state transition becomes saturated at surprisingly low powers.

We make several observations about this result. 
Firstly, the steady  state depends on the driving direction, which accounts for the asymmetric response to driving fields that has been discussed elsewhere.  
Secondly, the atomic steady state, $\rho_{ss}^{(\alpha)}$, is mixed, but retains some entanglement (with respect to any local atomic basis) between the atoms due to the dark state component: the concurrence of ${\bar \rho_{0}}$ is $C_{\bar \rho_{0}}=p_D^2$, so that $C_{\rho_{ss}^{(\alpha)}}\approx4/9>0$.
Lastly, $\rho_{ss}^{(\alpha)}$ exhibits steady-state population inversion, since the ground state population is $p_G\approx1/3<1/2$.  
 
For a symmetric system ($\delta=0$), the dark state is completely decoupled from the field and thus in principle infinitely long-lived.  
Conversely, for small $|\delta|>0$, the dark state is very weakly coupled to the waveguide, so that it has an anomalously long lifetime. 
As we show in appendix~\ref{app:symmbasis}, there is a fast time scale associated with the bright state \mbox{$\ket{B}=(\ket{ge}-\ket{eg})/\sqrt{2}$}, given by $\tau_B^{-1}=\gamma_B=2\gamma+O(\delta^2)$, 
and a slow time scale associated with the dark state, given by \mbox{$\tau_D^{-1}=\gamma_D=\delta^2\gamma+O(\delta^3)$}~\cite{RedcYuds14, FangBara17}. 
It is this slow timescale that leads to a saturation of the ground state to dark state transition at very low incident power. 

Finally, the purification \cite{nielsen_2000} of $\rho_{ss}^{(\alpha)}$ is the tripartite $\ket{W}$ state $$\ket{W}=(\ket{eg0}+\ket{ge0}+\ket{gg1})/\sqrt3,$$  
where we have introduced a purifying system labeled by the states $\ket{0}$ and $\ket{1}$. 
In this picture, $\ket{0}$ corresponds to the incident field being transmitted, and $\ket{1}$ corresponds to the incident field being reflected \cite{MakaLeto03}.
The evolution of the field-plus-atom is unitary, so that the purifying system has support on the field  modes $a_{\rm out}^\dag$ and $b_{\rm out}^\dag$ \cite{MakaLeto03,RedcYuds14}, corresponding to the `recently-scattered' field.  
That is, the slow subspace of the atoms is entangled with the field out to a distance $\sim\tau_D c_s$.  

\section{Adiabatic elimination \label{sec:Adiab}}
In light of the steady-state analysis above and the separation of time-scales discussed there, we adiabatically eliminate the fast subspace spanned by $\mathcal{F}=\{{\ket{B}},{\ket{E}}\}$, where $\ket{E}={{\ket{ee}}}$, to yield dynamics in the slow subspace, 
spanned by $\mathcal{S}=\{{\ket{G}},{\ket{D}}\}$. 
We apply the adiabatic elimination procedure from Refs.~\cite{BoutSilb08,BoutHandSilb08,CombKercSaro16}, to get the SLH triple for the system restricted to the slow subspace.  
As described in appendix~\ref{app:AdiabElim}, to lowest non-trivial order in $\alpha,\beta$ and $\delta$ we derive the adiabatically eliminated operators
\begin{align}
	\tilde H_T &= {\alpha\, \delta\,} \tilde\sigma_{x}/2 \,,\label{eq:HElim}\\
	\tilde L_{\rightarrow} &=  \tfrac12\big( (\alpha-\beta) \tilde\sigma_{z} -(\alpha+\beta)\mathds 1 \big) \,,\nn\\
	\tilde L_{\leftarrow} &=\ii \,\delta\, \tilde\sigma_{-}  - \tfrac12\big( (\alpha-\beta) \tilde\sigma_{z} +(\alpha+\beta)\mathds 1 \big)  \,,
	\label{eq:SLHElim}
\end{align}
where \mbox{$\tilde\sigma_{z} ={ \ket{G}}{\bra{G}} - {\ket{D}}{\bra{D}}$}, $\tilde\sigma_{x} = \tilde\sigma_{-} + \tilde \sigma_{+}$, and \mbox{$\tilde\sigma_{-} = \ket{G}\bra{D}$}.
We note that this adiabatic elimination does not rely on the earlier perturbative assumption that $\delta\ll\alpha,\beta$,  
rather it merely requires $\delta,\alpha,\beta\ll\gamma=1$.

The coherent part of the master equation generated by $\tilde H_T $ accounts for the asymmetry observed in the steady state: the effective Hamiltonian, $\tilde H_T$, vanishes for $\beta$ driving (i.e.\ from the right), so that the dynamics within the slow subspace is completely decoupled from $\beta$.  

\begin{figure}[t]
	\begin{center}
	\includegraphics[width=\columnwidth]{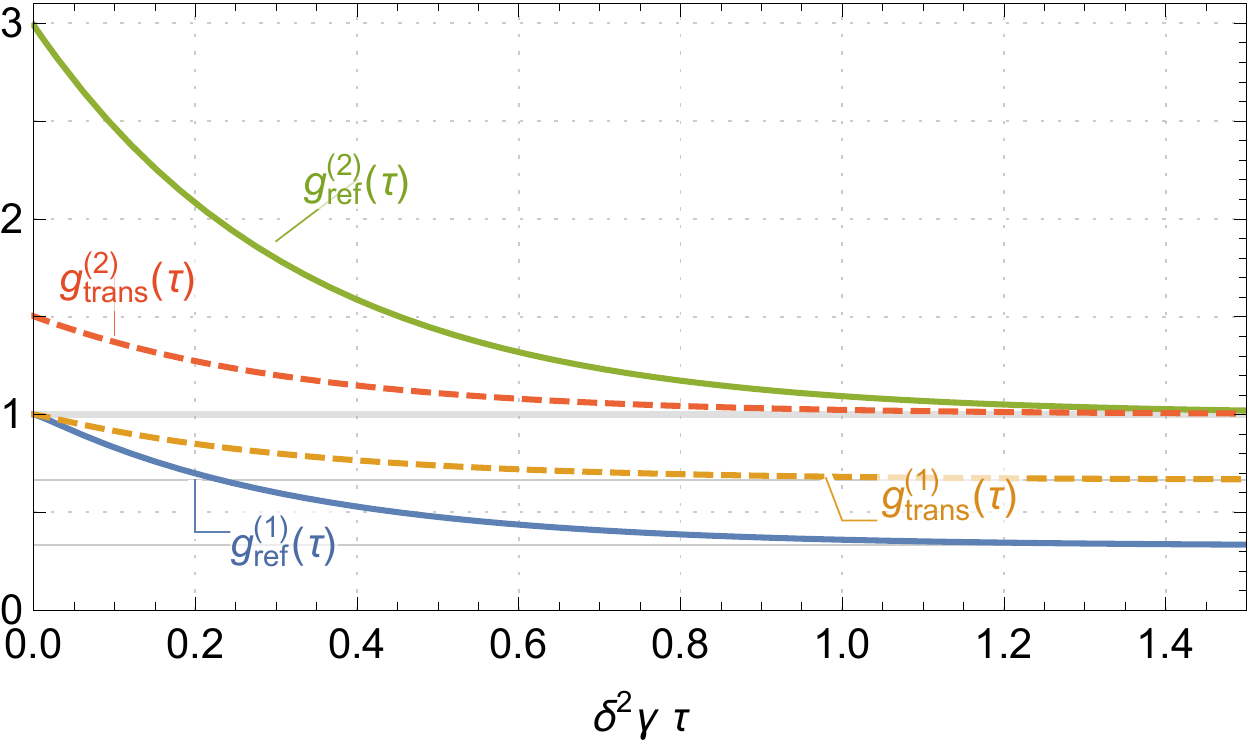}
	\caption{(colour online) Correlation functions, $g^{(1)}(\tau)$ and $g^{(2)}(\tau)$, for reflection and transmission, under $\alpha$ driving. 
		We can calculate these correlation functions using the full four-dimensional Hilbert space, the adiabatically-eliminated two-dimensional space, and the `flapping-mirror' toy-model.  
		The differences between these calculation methods are $\sim(\delta/\alpha)^2=10^{-6}$, so are not visible on this scale.
	}
	\label{fig:g12}
	\end{center}
\end{figure}

The effective Lindblad operator $\tilde L_{\leftarrow}$ is a coherent combination of dephasing and relaxation. 
Together with the $\alpha$-dependent driving in $\tilde H_T$,  the system evolves to an inverted  steady state: without the interplay between driving and dissipation, 
the maximum population of the dark state would be bounded by $p_D<1/2$ \cite{stace2005population, Hughes:PRL:2011, Hughes:NJP:2013, stace2013dynamical, Colless:2014aa}.

The steady-state in the slow subspace, $\mathcal{S}$, is then
\begin{align}
	\tilde\rho_{ss}^{(\alpha,\beta,\delta)} \!=\! \frac{1}{6\alpha^{2} \!+\!2\beta^{2} \!+\!\delta^{2}}\!\l[ \begin{array}{cc}
			2\alpha^{2} \!+\! 2\beta^{2} \!+\! \delta^{2} \quad& 2\ii \alpha\delta\! \\
			-2\ii \alpha \delta & 4\alpha^{2}\!
		\end{array} \r].\label{eqn:reducedMEsol}
\end{align}
This result does not require $\delta\ll |\alpha|+|\beta|$ as in \eqns{eq:pDG} and \eqref{eqn:rhoss}, however, naturally, it agrees with those results in the limits $\alpha\rightarrow0$ or $\beta\rightarrow0$. 
Together with $\tilde L_{\rightleftharpoons}$ in Eq.~\eqref{eq:SLHElim}, \eqn{eqn:reducedMEsol} enables us to find analytical expressions for the field fluxes.  

\section{Scattering \label{sec:Scatter}} 
Using the adiabatically eliminated operators, we  calculate the scattering matrices for field amplitudes, $S$, and fluxes, $T$.  Writing
\begin{align}
	\l[ \begin{array}{c}
		\alpha_{\text{out}} \\
		\beta_{\text{out}}
	\end{array} \r] = S \l[ \begin{array}{c}
		\alpha
		 \\
		\beta
	\end{array} \r] \,, \quad
	\l[ \begin{array}{c}
		\mathcal{A}_{\text{out}} \\
		\mathcal{B}_{\text{out}}
	\end{array} \r] \!= T \l[ \begin{array}{c}
		{\mathcal{A}_{\text{in}}} \\
		{\mathcal{B}_{\text{in}}}
	\end{array} \r] \,,\nn
\end{align}
where 
${\mathcal{A}_{\text{in}}}=|\alpha|^2$, ${\mathcal{B}_{\text{in}}}=|\beta|^2$, we find
\begin{align}
	S &= -\l[ \begin{array}{cc}
		\frac{2\alpha^{2} - \delta^{2}}{6\alpha^{2} +\delta^{2}} & 0 \\
		\frac{4\alpha^{2}}{6\alpha^{2} + \delta^{2}} & 1
	\end{array} \r] \approx -\l[ \begin{array}{cc}
		1/3 & 0 \\
		2/3 & 1
	\end{array} \r]
	,\label{eqn:S}\\
	T &\!=\!\l[ \begin{array}{cc}
		\!\!\mathcal{R}_\alpha\! &\! \mathcal{T}_\beta\!\! \\
		\!\!\mathcal{T}_\alpha\! &\! \mathcal{R}_\beta\!\!
	\end{array} \r]
	\!=\!
	 \l[ \begin{array}{cc}
		\!p_G^{(\alpha,0,\delta)} &\! 	p_D^{(0,\beta,\delta)} \!\\
		\!p_D^{(\alpha,0,\delta)} \!& \! p_G^{(0,\beta,\delta)}\!
	\end{array} \r] 
	\!\approx\! \l[ \begin{array}{cc}
		\!1/3 &\! 0\! \\
		\!2/3 &\! 1\!
	\end{array} \r] \!
	,\label{eqn:T}
\end{align}
where $p_K^{(\alpha,\beta,\delta)}=\bra{K}\tilde\rho_{ss}^{(\alpha,\beta,\delta)}\ket{K}$, and the approximations hold for $\delta\ll|\alpha|+|\beta|$. Up to an overall sign, the scattering matrices in \eqns{eqn:S} and (\ref{eqn:T}) are identical.  

We see that $S\neq S^T$, consistent with the definition of an (imperfect) isolator.  For this system, the Lorentz reciprocity theorem is broken by the nonlinear saturation of the atoms \cite{JalaPetrEich13}.

Further, $S$ is not unitary, indicating that the scattered field is not fully coherent for $\alpha$ driving.
One way to see this is to note that for $\alpha\neq0$ we find $\mathcal{A}_{\text{out}}\neq|\alpha_{\text{out}}|^2$,
i.e.\ the output field flux is not equal to the square of the output field amplitude, as it would be for a coherent state. 
Conversely, for $\beta$ driving,  the dark state remains unpopulated, and if $|\beta_{\text{in}}|^2\ll\gamma$, the bright state will be unsaturated, so that the atoms will  reflect the incident field \cite{1367-2630-15-3-035009}.  
In this case, the output field is  coherent as it is simply the reflected input field.

While these observations imply the field is incoherent, there are many different ways in which incoherence may be manifest.  To  quantify the incoherent scattering for $\alpha$ driving, we calculate the steady-state output-field correlation functions,
\begin{eqnarray}
g_\text{ref}^{(1)}(\tau)&=&\langle a_{\text{out}}^\dagger(t+\tau) a_{\text{out}}(t) \rangle/\mathcal{A}_{\text{out}},\nonumber\\
g^{(2)}_\text{ref}(\tau)&=&\langle a_{\text{out}}^\dagger(t) a_{\text{out}}^\dagger(t+\tau) a_{\text{out}}(t+\tau)a_{\text{out}}(t) \rangle/\mathcal{A}_{\text{out}}^2\nonumber
\end{eqnarray} 
 and similar for the transmitted field correlations, $g_\text{trans}^{(1,2)}(\tau)$~\cite{FangBara17} (see also appendix~\ref{app:CorrFunc}).  Further, the spectrum of the scattered field can be computed directly from $g^{(2)}_\text{ref}(\tau)$, so this may be useful for experimental comparison.  We will later compare these correlation functions to the result of a simple ``flapping mirror" model, and show that they are essentially indistinguishable.

For $\alpha$ driving (from the dark state coupling direction), the two-time field-field correlations functions, $g^{(1)}(\tau)$, for the reflected and transmitted fields satisfy \mbox{$g^{(1)}_{\text{ref}} (0)=g^{(1)}_{\text{trans}} (0)=1$}, and
\begin{align}
	\lim_{\tau\rightarrow\infty} g^{(1)}_{\text{ref}}(\tau) &= \frac{(2\alpha^{2} - \delta^{2})^{2}}{(4\alpha^{2} +\delta^{2})^{2} - 4\alpha^{4}} = \!\frac13+O\Big(\frac{\delta^2}{\alpha^2}\!\Big),\nn\\
	\lim_{\tau\rightarrow\infty}  g^{(1)}_{\text{trans}}(\tau) &= \frac{4\alpha^{2}}{6\alpha^{2} + \delta^{2}} = \frac23+O\Big(\frac{\delta^2}{\alpha^2}\Big) \,.\nn
\end{align}

The flux-flux correlation function $g^{(2)}$ satisfies \mbox{$\lim_{\tau\rightarrow\infty}g^{(2)}_{\text{ref}} (\tau)=\lim_{\tau\rightarrow\infty}g^{(1)}_{\text{trans}} (\tau)=1$}, and
\begin{align}
	g^{(2)}_{\text{ref}}(0)& = \frac{6\alpha^{2} + \delta^{2}}{2\alpha^{2}+\delta^{2}} = 3+O\Big(\frac{\delta^2}{\alpha^2}\Big) ,\nn\\
	g^{(2)}_{\text{trans}}(0)& = \frac32 +\frac{\delta^{2}}{4\alpha^{2}} .\nn
\end{align}
At long times, $g^{(1)}(\tau)$ is sub-unity, indicating incoherent statistics, while $g^{(2)}(\tau)>1$, indicating thermal or bunched light.   
At intermediate times, the correlation functions decay exponentially between the above limits, as shown in Fig.~\ref{fig:g12}. 

The correlation functions for $\beta$ driving (from the decoupled direction) are unity for all time (up to corrections of order $\delta^2/\alpha^2$).  

The incoherence of the outgoing field under $\alpha$ driving arises from two competing effects: firstly, the drive couples weakly to the dark state, so that when \mbox{$\delta^2\gamma\ll\alpha^2\ll\gamma=1$}, the dark state becomes ultra-saturated (i.e.\ inverted) over a time $\sim \tau_D$.  
Secondly, the input field reflects off the strongly coupled dipole transition  $\ket{G}\leftrightarrow\ket{B}$, so that when the system is shelved in the dark state, $\ket{D}$, it becomes transparent.  
At steady-state, the system thus fluctuates between the ground state, which coherently reflects the incoming field as would a single-atom mirror \cite{1367-2630-15-3-035009,PhysRevA.89.013805}, and the dark state, which is transparent to the driving field.  
The normalised output flux is thus equal to the dark state probability, $p_D=2/3$.

\begin{figure}[t]
	\begin{center}
	\includegraphics[width=\columnwidth]{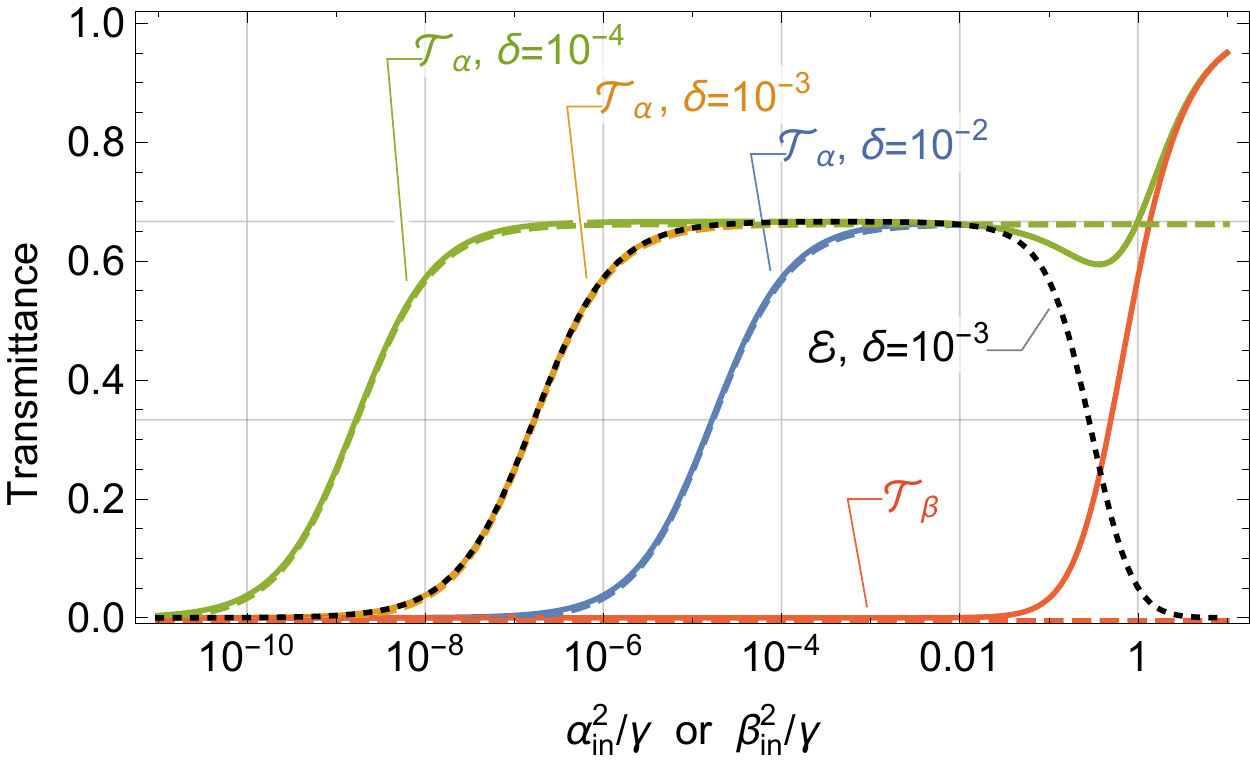}
	\caption{(colour online) Transmittance versus incident power, $|\alpha_{\text{in}}|^2/\gamma$ or $|\beta_{\text{in}}|^2/\gamma$.  
		The dark state becomes saturated for $|\alpha|^2\gtrsim\delta^2$, so that $\mathcal{T}_\alpha$  has a dark-state saturation plateau for $\delta^2\lesssim|\alpha|^2/\gamma\lesssim1$ wherein $\mathcal{T}_\alpha\approx2/3=p_D$.   
		$\mathcal{T}_\beta$  is independent of $\delta$.  
		The bright state also becomes saturated when $|\alpha|^2,|\beta|^2\gtrsim\gamma$, so that $\mathcal{T}_{\alpha}$ and $\mathcal{T}_{\beta}$ both asymptote to unity when $|\alpha|^2,|\beta|^2\gtrsim\gamma$.  
		Solid curves are calculated using the full four-dimensional Hilbert space. Dashed curves are calculated for the two-dimensional slow subspace (adiabatically elimination) which are valid for $|\alpha|^2,|\beta|^2\ll\gamma$, 
		and are plotted with a small  offset for visibility.  
		Also shown is the `diode efficiency' $\mathcal{E}$ (dotted)  for $\delta=10^{-3}$, as defined in \cite{DaiRoulLe15}.
	}
	\label{fig:T}
	\end{center}
\end{figure}

\section{Poisson Rate Equations and Flapping Mirror Model \label{sec:FM}}
The correlation functions can be understood from a simple rate model in which we eliminate the off-diagonal elements $\rho_{DG}$ and $\rho_{GD}$ in \eqn{eqn:lk}, 
assuming the reduced Hamiltonian and Lindblad operators given in \eqn{eq:SLHElim}. As described in Appendix \ref{app:rate}, for  $\alpha$-driving, we find\
\begin{align}
	\l[ \begin{array}{c}
		\dot P_D \\
		\dot P_G
	\end{array} \r] &\approx\left[
	\begin{array}{cc}
		 -\delta^2 &  \hphantom{-}2 \delta^2\\
		 \hphantom{-}\delta^2 & -2 \delta^2 \\
	\end{array}
	\right] \l[ \begin{array}{c}
		P_D \\
		P_G
	\end{array} \r]. 
	\label{eqn:rate}
\end{align}

Since the dark state, $\ket{D}$, is transparent to the incident field (so the reflectivity of the system is $R=0$), and the ground state $\ket{G}$ reflects the field (so the reflectivity is $R=1$),  this expression motivates a simple ``flapping-mirror"  classical rate model,  which  replicates the output field correlation functions. 

Suppose a black-box optical-circuit consists of  a mirror which flips in  and out  of the optical path, controlled by a two-state random variable $R\in\{0,1\}\leftrightarrow\{D,G\}$, where the arrow indicates a precise correspondence between the notional reflectivity of the mirror, and the state of the two-atom system in the reduced slow subspace. 
For $R=1$, the  optical circuit is fully reflective (i.e.\ reflectance $\mathcal R_1=1$), and when $R=0$ the circuit is fully transparent, (i.e.\ reflectance $\mathcal R_0=0$). 
We assume the black box responds asymmetrically  to light incident from different directions
\footnote{Such a scheme is physically realisable by using a {QND} detection of photon flux to determine from which direction the field is incident, and controlling the mirror accordingly}:
for light incident from the right, we fix $R=1$ so that $\mathcal R_1=1$; for light incident from the left we drive the state of the mirror 
with a Poisson process following a simple two-state rate model with transition  rates $\Gamma_{R \overline R}$, from state $R$ to $\overline R$, given by the matrix elements in \eqn{eqn:rate}.
Starting in state $R$, the flapping mirror will be found in that state after time $\tau$  with probability 
\begin{equation}
P_{R,R}(\tau)=p_{R}-(1-p_{R})e^{-\Gamma_{\text{tot}} t}
\end{equation}
where \mbox{$p_0=p_D=2/3$} and \mbox{$p_1=p_G=1/3$} are the steady-state probabilities of state $R$, i.e.\ \mbox{$p_{R}=\Gamma_{\overline{R} R}/\Gamma_{\text{tot}}$}, \mbox{$\Gamma_{01}=\Gamma_D=\delta^2\gamma=\Gamma_{10}/2$},  and $\Gamma_{\text{tot}}=\Gamma_{0 1}+\Gamma_{1 0}=3\delta^2\gamma$.  

In appendix~\ref{app:Flap} we discuss in detail the statistics of a coherent state of amplitude $\alpha_{\text{in}}$  passing through this flapping mirror device.  
The reflected and transmitted field {amplitudes} are \mbox{$\{\alpha_{\text{out}},\beta_{\text{out}}\}=\{p_1\alpha,p_0\alpha\}$},  
the fluxes  are $\{\mathcal{A}_{\text{out}},\mathcal{B}_{\text{out}}\}=\{p_1 |\alpha|^2,p_0 |\alpha|^2\}$, 
and the correlation functions are 
\begin{eqnarray}
g_{\text{ref}}^{(1)}(\tau)&=&P_{1,1}(\tau),\\
g_{\text{trans}}^{(1)}(\tau)&=&P_{0,0}(\tau),\\
g_{\text{ref}}^{(2)}(\tau)&=&P_{1,1}(\tau)/p_1,\\
g_{\text{trans}}^{(2)}(\tau)&=&P_{0,0}(\tau)/p_0.
\end{eqnarray}


We see that the output field amplitudes and fluxes of the flapping mirror model agree with the  two-atom scattering results in \eqns{eqn:S} and (\ref{eqn:T}) (up to overall phase). The correlation functions 
of the   flapping mirror model completely replicate the corresponding correlation functions of the $\alpha$-driven, two-atom system, and are visually indistinguishable from the traces plotted in \fig{fig:g12}.

\section{Rectification \label{sec:Rect}}
The asymmetry of the flux through the device has attracted some interest due to an analogy with a diode: the system appears to `rectify' flux from one side.  
This is manifest in the transmission coefficients, in \fig{fig:T}, which shows $\mathcal{T}_\alpha$ for several different values of $\delta$, and $\mathcal{T}_\beta$, which is essentially independent of $\delta$.  
There is a clear plateau  where $\mathcal{T}_\alpha\approx2/3$, for $\delta^2\gamma\ll\alpha_{\text{in}}\ll\gamma=1$.  This is consistent with the flux scattering matrix elements in \eqn{eqn:T}, which shows that the transmission coefficient, $\mathcal{T}_K$ is just the dark state probability.  
The low-power roll-off of $\mathcal{T}_\alpha$  corresponds to the saturation intensity of the dark state, which is small due to the extremely weak dark-state coupling at small $\delta$.

Previous work has quantified this asymmetry using   the `diode efficiency' \cite{DaiRoulLe15}, defined as
$$\mathcal{E}=\mathcal{T}_\alpha(\mathcal{T}_\alpha-\mathcal{T}_\beta)/(\mathcal{T}_\alpha+\mathcal{T}_\beta).$$  
Clearly, in the regime where the asymmetry is greatest, $\mathcal{T}_\beta\ll\mathcal{T}_\alpha$, so that $\mathcal{E}\approx\mathcal{T}_\alpha=p_D(\alpha,0,\delta)$.  
The diode efficiency $\mathcal{E}$ is shown as a dotted curve in \fig{fig:T}, and tracks $\mathcal{T}_\alpha$ until the bright state starts to become saturated at high power.  
As discussed in appendix~\ref{app:optimum} we numerically optimise $\mathcal{E}$ over $\delta\phi$ and $\delta\omega$, and we find the maximum value, $\mathcal{E}=p_D\approx2/3$, at the settings $\delta\phi=-\delta\omega=\delta$ that we have adopted throughout.

\begin{figure}[t]
	\begin{center}
	\includegraphics[width=\columnwidth]{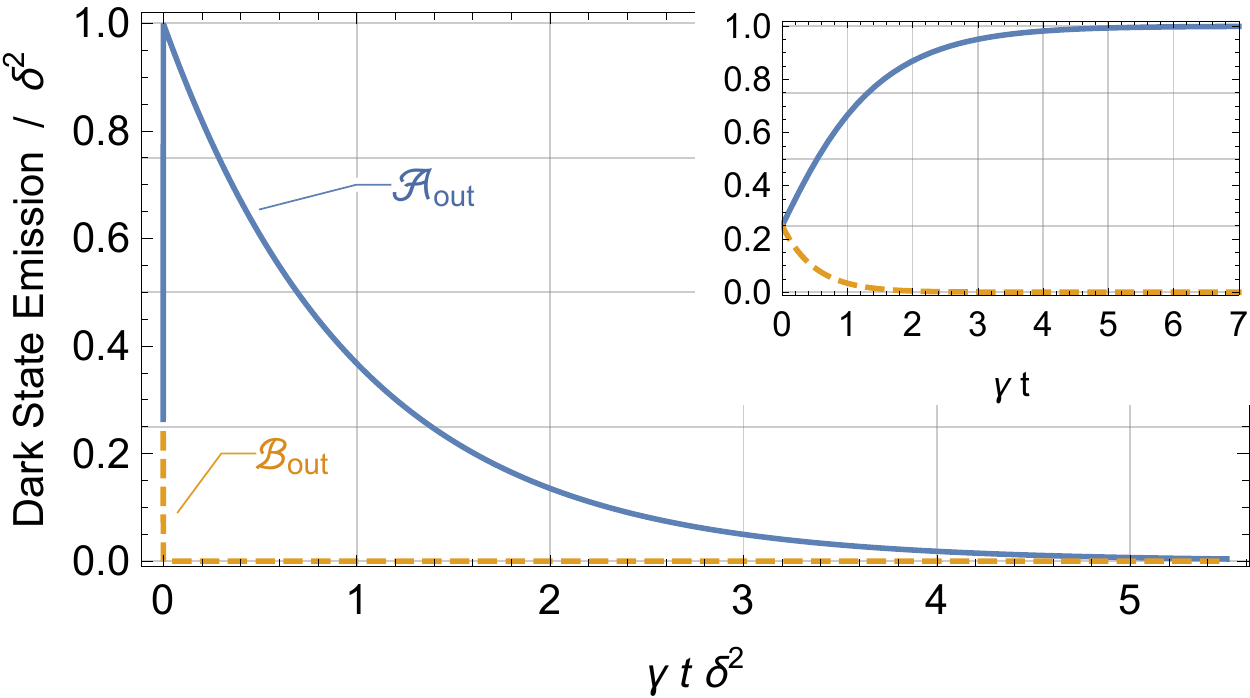}
	\caption{(colour online) Radiated flux from the dark state, $\ket{D}$, moving left, $\mathcal{A}_\text{out}$, and  right, $\mathcal{B}_\text{out}$, showing strong asymmetry in the emission profile, analogous to a Yagi-Uda directional transmitter. 
		The main panel shows flux over long time-scales $\gamma t\sim \delta^{-2}\gg1$; inset shows flux over short time-scales, $\gamma t\sim1$.
	}
	\label{fig:asymm}
	\end{center}
\end{figure}

The underlying asymmetry in absorption is manifest from the asymmetric field coupling in the effective Hamiltonian, \eqn{eq:HElim}, and this also gives rise to asymmetric emission from atomic excited states.  \fig{fig:asymm} shows the time-dependent, left-going flux, $\mathcal{A}_\textrm{out}$, and  right-going flux, $\mathcal{B}_\textrm{out}$, from the dark state \cite{MakaLeto03}. It is evident that the dark state emits asymmetrically, with the vast majority of energy radiating to the left over a long time $\gamma t\sim\delta^{-2}$.   This asymmetric emission is reminiscent of a two-element Yagi-Uda  antenna (of the kind used for directional radio transceivers), in which detuned dipole elements behave as `reflectors' and `directors' to produce a directional radiation pattern \cite{Kosako:2010aa,LembRsheAldo13}. 

For $n=2$ atoms, the peak ``diode efficiency" \mbox{$\mathcal{E}=2/3=n/(n+1)$} is equal to the dark state population in the entangled tripartite $\ket{W}$ state of the atoms and field.  By analogy, we speculate that if $n>2$ atoms are suitably tuned (e.g.\ as in Refs.~\cite{StanRablZoll12,RamoPichDale14,PichRamoDale15}), then the  diode efficiency $\mathcal{E}$ could be improved for larger  $n$, albeit  with a narrow bandwidth.

\section{Conclusions}
To conclude, we have analysed a two-atom system in a 1D waveguide.  
Consistent with previous results, we find that this system responds asymmetrically to external driving fields when one of the atoms is detuned from the driving field and their separation is close to a half-integer multiple of the wavelength, 
and we establish the regime where the response is maximally asymmetric.

In addition, we have shown that the onset time for the asymmetry is not instantaneous.  Rather, it is established over a time scale set by the dark-state lifetime,  which sets a modulation bandwidth for any time-varing signals. 
The asymmetry in the field coupling leads to an inverted, entangled mixture of the ground and dark states, which is ultimately responsible for nonreciprocal scattering. The scattered field statistics are replicated by a simple stochastic flapping-mirror model, which physically corresponds to fluctuations between the dark and ground states of the atoms.

\acknowledgements{We thank M.~Pletyukhov and P.~Rabl for stimulating  discussions. 
This work was supported by the Australian Research Council under the Discovery and Centre of Excellence funding schemes (project numbers DP150101033, DE160100356, and CE110001013).
}

\bibstyle{iopart-num}
\bibliography{MicroDiode}

\begin{thebibliography}{33}%
\makeatletter
\providecommand \@ifxundefined [1]{%
 \@ifx{#1\undefined}
}%
\providecommand \@ifnum [1]{%
 \ifnum #1\expandafter \@firstoftwo
 \else \expandafter \@secondoftwo
 \fi
}%
\providecommand \@ifx [1]{%
 \ifx #1\expandafter \@firstoftwo
 \else \expandafter \@secondoftwo
 \fi
}%
\providecommand \natexlab [1]{#1}%
\providecommand \enquote  [1]{``#1''}%
\providecommand \bibnamefont  [1]{#1}%
\providecommand \bibfnamefont [1]{#1}%
\providecommand \citenamefont [1]{#1}%
\providecommand \href@noop [0]{\@secondoftwo}%
\providecommand \href [0]{\begingroup \@sanitize@url \@href}%
\providecommand \@href[1]{\@@startlink{#1}\@@href}%
\providecommand \@@href[1]{\endgroup#1\@@endlink}%
\providecommand \@sanitize@url [0]{\catcode `\\12\catcode `\$12\catcode
  `\&12\catcode `\#12\catcode `\^12\catcode `\_12\catcode `\%12\relax}%
\providecommand \@@startlink[1]{}%
\providecommand \@@endlink[0]{}%
\providecommand \url  [0]{\begingroup\@sanitize@url \@url }%
\providecommand \@url [1]{\endgroup\@href {#1}{\urlprefix }}%
\providecommand \urlprefix  [0]{URL }%
\providecommand \Eprint [0]{\href }%
\providecommand \doibase [0]{http://dx.doi.org/}%
\providecommand \selectlanguage [0]{\@gobble}%
\providecommand \bibinfo  [0]{\@secondoftwo}%
\providecommand \bibfield  [0]{\@secondoftwo}%
\providecommand \translation [1]{[#1]}%
\providecommand \BibitemOpen [0]{}%
\providecommand \bibitemStop [0]{}%
\providecommand \bibitemNoStop [0]{.\EOS\space}%
\providecommand \EOS [0]{\spacefactor3000\relax}%
\providecommand \BibitemShut  [1]{\csname bibitem#1\endcsname}%
\let\auto@bib@innerbib\@empty
\bibitem [{\citenamefont {Fratini}\ \emph {et~al.}(2014)\citenamefont
  {Fratini}, \citenamefont {Mascarenhas}, \citenamefont {Safari}, \citenamefont
  {Poizat}, \citenamefont {Valente}, \citenamefont {Auff\`eves}, \citenamefont
  {Gerace},\ and\ \citenamefont {Santos}}]{FratMascSafa14}%
  \BibitemOpen
  \bibfield  {author} {\bibinfo {author} {\bibfnamefont {F.}~\bibnamefont
  {Fratini}}, \bibinfo {author} {\bibfnamefont {E.}~\bibnamefont
  {Mascarenhas}}, \bibinfo {author} {\bibfnamefont {L.}~\bibnamefont {Safari}},
  \bibinfo {author} {\bibfnamefont {J.-P.}\ \bibnamefont {Poizat}}, \bibinfo
  {author} {\bibfnamefont {D.}~\bibnamefont {Valente}}, \bibinfo {author}
  {\bibfnamefont {A.}~\bibnamefont {Auff\`eves}}, \bibinfo {author}
  {\bibfnamefont {D.}~\bibnamefont {Gerace}}, \ and\ \bibinfo {author}
  {\bibfnamefont {M.~F.}\ \bibnamefont {Santos}},\ }\bibfield  {title}
  {\enquote {\bibinfo {title} {Fabry-perot interferometer with quantum mirrors:
  Nonlinear light transport and rectification},}\ }\href
  {https://dx.doi.org/10.1103/PhysRevLett.113.243601} {\bibfield  {journal}
  {\bibinfo  {journal} {Phys. Rev. Lett.}\ }\textbf {\bibinfo {volume} {113}},\
  \bibinfo {pages} {243601} (\bibinfo {year} {2014})}\BibitemShut {NoStop}%
\bibitem [{\citenamefont {Dai}\ \emph {et~al.}(2015)\citenamefont {Dai},
  \citenamefont {Roulet}, \citenamefont {Le},\ and\ \citenamefont
  {Scarani}}]{DaiRoulLe15}%
  \BibitemOpen
  \bibfield  {author} {\bibinfo {author} {\bibfnamefont {J.}~\bibnamefont
  {Dai}}, \bibinfo {author} {\bibfnamefont {A.}~\bibnamefont {Roulet}},
  \bibinfo {author} {\bibfnamefont {H.~N.}\ \bibnamefont {Le}}, \ and\ \bibinfo
  {author} {\bibfnamefont {V.}~\bibnamefont {Scarani}},\ }\bibfield  {title}
  {\enquote {\bibinfo {title} {Rectification of light in the quantum regime},}\
  }\href {https://dx.doi.org/10.1103/PhysRevA.92.063848} {\bibfield  {journal}
  {\bibinfo  {journal} {Phys. Rev. A}\ }\textbf {\bibinfo {volume} {92}},\
  \bibinfo {pages} {063848} (\bibinfo {year} {2015})}\BibitemShut {NoStop}%
\bibitem [{\citenamefont {Mascarenhas}\ \emph {et~al.}(2016)\citenamefont
  {Mascarenhas}, \citenamefont {Santos}, \citenamefont {Auff\`eves},\ and\
  \citenamefont {Gerace}}]{MascSantAuff16}%
  \BibitemOpen
  \bibfield  {author} {\bibinfo {author} {\bibfnamefont {E.}~\bibnamefont
  {Mascarenhas}}, \bibinfo {author} {\bibfnamefont {M.~F.}\ \bibnamefont
  {Santos}}, \bibinfo {author} {\bibfnamefont {A.}~\bibnamefont {Auff\`eves}},
  \ and\ \bibinfo {author} {\bibfnamefont {D.}~\bibnamefont {Gerace}},\
  }\bibfield  {title} {\enquote {\bibinfo {title} {Quantum rectifier in a
  one-dimensional photonic channel},}\ }\href
  {https://dx.doi.org/10.1103/PhysRevA.93.043821} {\bibfield  {journal}
  {\bibinfo  {journal} {Phys. Rev. A}\ }\textbf {\bibinfo {volume} {93}},\
  \bibinfo {pages} {043821} (\bibinfo {year} {2016})}\BibitemShut {NoStop}%
\bibitem [{\citenamefont {Fratini}\ and\ \citenamefont
  {Ghobadi}(2016)}]{FratGhob16}%
  \BibitemOpen
  \bibfield  {author} {\bibinfo {author} {\bibfnamefont {F.}~\bibnamefont
  {Fratini}}\ and\ \bibinfo {author} {\bibfnamefont {R.}~\bibnamefont
  {Ghobadi}},\ }\bibfield  {title} {\enquote {\bibinfo {title} {{Full quantum
  treatment of a light diode}},}\ }\href
  {https://dx.doi.org/10.1103/PhysRevA.93.023818} {\bibfield  {journal}
  {\bibinfo  {journal} {Physical Review A}\ }\textbf {\bibinfo {volume} {93}},\
  \bibinfo {pages} {023818} (\bibinfo {year} {2016})}\BibitemShut {NoStop}%
\bibitem [{\citenamefont {Gonzalez-Ballestero}\ \emph
  {et~al.}(2016)\citenamefont {Gonzalez-Ballestero}, \citenamefont {Moreno},
  \citenamefont {Garcia-Vidal},\ and\ \citenamefont
  {Gonzalez-Tudela}}]{GonzMoreGarc16}%
  \BibitemOpen
  \bibfield  {author} {\bibinfo {author} {\bibfnamefont {C.}~\bibnamefont
  {Gonzalez-Ballestero}}, \bibinfo {author} {\bibfnamefont {E.}~\bibnamefont
  {Moreno}}, \bibinfo {author} {\bibfnamefont {F.~J.}\ \bibnamefont
  {Garcia-Vidal}}, \ and\ \bibinfo {author} {\bibfnamefont {A.}~\bibnamefont
  {Gonzalez-Tudela}},\ }\bibfield  {title} {\enquote {\bibinfo {title}
  {Nonreciprocal few-photon routing schemes based on chiral waveguide-emitter
  couplings},}\ }\href {https://dx.doi.org/10.1103/PhysRevA.94.063817}
  {\bibfield  {journal} {\bibinfo  {journal} {Phys. Rev. A}\ }\textbf {\bibinfo
  {volume} {94}},\ \bibinfo {pages} {063817} (\bibinfo {year}
  {2016})}\BibitemShut {NoStop}%
\bibitem [{\citenamefont {Ordonez-Miranda}\ \emph {et~al.}(2017)\citenamefont
  {Ordonez-Miranda}, \citenamefont {Ezzahri},\ and\ \citenamefont
  {Joulain}}]{OrdoEzzaJoul17}%
  \BibitemOpen
  \bibfield  {author} {\bibinfo {author} {\bibfnamefont {J.}~\bibnamefont
  {Ordonez-Miranda}}, \bibinfo {author} {\bibfnamefont {Y.}~\bibnamefont
  {Ezzahri}}, \ and\ \bibinfo {author} {\bibfnamefont {K.}~\bibnamefont
  {Joulain}},\ }\bibfield  {title} {\enquote {\bibinfo {title} {Quantum thermal
  diode based on two interacting spinlike systems under different
  excitations},}\ }\href {https://dx.doi.org/10.1103/PhysRevE.95.022128}
  {\bibfield  {journal} {\bibinfo  {journal} {Phys. Rev. E}\ }\textbf {\bibinfo
  {volume} {95}},\ \bibinfo {pages} {022128} (\bibinfo {year}
  {2017})}\BibitemShut {NoStop}%
\bibitem [{\citenamefont {Fang}\ and\ \citenamefont
  {Baranger}(2017)}]{FangBara17}%
  \BibitemOpen
  \bibfield  {author} {\bibinfo {author} {\bibfnamefont {Y.-L.~L.}\
  \bibnamefont {Fang}}\ and\ \bibinfo {author} {\bibfnamefont {H.~U.}\
  \bibnamefont {Baranger}},\ }\bibfield  {title} {\enquote {\bibinfo {title}
  {Cascaded emitters in a waveguide: Non-reciprocity and correlated photons at
  perfect elastic transmission},}\ }\href
  {https://doi.org/10.1103/PhysRevA.96.013842} {\bibfield  {journal} {\bibinfo
  {journal} {Physical Review A}\ }\textbf {\bibinfo {volume} {96}} (\bibinfo
  {year} {2017})}\BibitemShut {NoStop}%
\bibitem [{\citenamefont {Rudolph}\ \emph {et~al.}(1995)\citenamefont
  {Rudolph}, \citenamefont {Ficek},\ and\ \citenamefont
  {Dalton}}]{RudoFiceDalt95}%
  \BibitemOpen
  \bibfield  {author} {\bibinfo {author} {\bibfnamefont {T.~G.}\ \bibnamefont
  {Rudolph}}, \bibinfo {author} {\bibfnamefont {Z.}~\bibnamefont {Ficek}}, \
  and\ \bibinfo {author} {\bibfnamefont {B.~J.}\ \bibnamefont {Dalton}},\
  }\bibfield  {title} {\enquote {\bibinfo {title} {Two-atom resonance
  fluorescence in running- and standing-wave laser fields},}\ }\href
  {https://dx.doi.org/10.1103/PhysRevA.52.636} {\bibfield  {journal} {\bibinfo
  {journal} {Phys. Rev. A}\ }\textbf {\bibinfo {volume} {52}},\ \bibinfo
  {pages} {636} (\bibinfo {year} {1995})}\BibitemShut {NoStop}%
\bibitem [{\citenamefont {Rudolph}\ and\ \citenamefont
  {Ficek}(1998)}]{RudoFice98}%
  \BibitemOpen
  \bibfield  {author} {\bibinfo {author} {\bibfnamefont {T.}~\bibnamefont
  {Rudolph}}\ and\ \bibinfo {author} {\bibfnamefont {Z.}~\bibnamefont
  {Ficek}},\ }\bibfield  {title} {\enquote {\bibinfo {title} {Interference
  pattern with a dark center from two atoms driven by a coherent laser
  field},}\ }\href {\doibase 10.1103/PhysRevA.58.748} {\bibfield  {journal}
  {\bibinfo  {journal} {Phys. Rev. A}\ }\textbf {\bibinfo {volume} {58}},\
  \bibinfo {pages} {748} (\bibinfo {year} {1998})}\BibitemShut {NoStop}%
\bibitem [{\citenamefont {Makarov}\ and\ \citenamefont
  {Letokhov}(2003)}]{MakaLeto03}%
  \BibitemOpen
  \bibfield  {author} {\bibinfo {author} {\bibfnamefont {A.~A.}\ \bibnamefont
  {Makarov}}\ and\ \bibinfo {author} {\bibfnamefont {V.~S.}\ \bibnamefont
  {Letokhov}},\ }\bibfield  {title} {\enquote {\bibinfo {title} {Spontaneous
  decay in a system of two spatially spearated atoms (one-dimensional case)},}\
  }\href {https://dx.doi.org/10.1134/1.1625059} {\bibfield  {journal} {\bibinfo
   {journal} {Journal of Experimental and Theoretical Physics}\ }\textbf
  {\bibinfo {volume} {97}},\ \bibinfo {pages} {688} (\bibinfo {year}
  {2003})}\BibitemShut {NoStop}%
\bibitem [{\citenamefont {Lembessis}\ \emph {et~al.}(2013)\citenamefont
  {Lembessis}, \citenamefont {Rsheed}, \citenamefont {Aldossary},\ and\
  \citenamefont {Ficek}}]{LembRsheAldo13}%
  \BibitemOpen
  \bibfield  {author} {\bibinfo {author} {\bibfnamefont {V.~E.}\ \bibnamefont
  {Lembessis}}, \bibinfo {author} {\bibfnamefont {A.~A.}\ \bibnamefont
  {Rsheed}}, \bibinfo {author} {\bibfnamefont {O.~M.}\ \bibnamefont
  {Aldossary}}, \ and\ \bibinfo {author} {\bibfnamefont {Z.}~\bibnamefont
  {Ficek}},\ }\bibfield  {title} {\enquote {\bibinfo {title} {Two-atom system
  as a nanoantenna for mode switching and light routing},}\ }\href
  {https://dx.doi.org/10.1103/PhysRevA.88.053814} {\bibfield  {journal}
  {\bibinfo  {journal} {Phys. Rev. A}\ }\textbf {\bibinfo {volume} {88}},\
  \bibinfo {pages} {053814} (\bibinfo {year} {2013})}\BibitemShut {NoStop}%
\bibitem [{\citenamefont {Laakso}\ and\ \citenamefont
  {Pletyukhov}(2014)}]{LaakPlet14}%
  \BibitemOpen
  \bibfield  {author} {\bibinfo {author} {\bibfnamefont {M.}~\bibnamefont
  {Laakso}}\ and\ \bibinfo {author} {\bibfnamefont {M.}~\bibnamefont
  {Pletyukhov}},\ }\bibfield  {title} {\enquote {\bibinfo {title} {Scattering
  of two photons from two distant qubits: Exact solution},}\ }\href
  {https://dx.doi.org/10.1103/PhysRevLett.113.183601} {\bibfield  {journal}
  {\bibinfo  {journal} {Phys. Rev. Lett.}\ }\textbf {\bibinfo {volume} {113}},\
  \bibinfo {pages} {183601} (\bibinfo {year} {2014})}\BibitemShut {NoStop}%
\bibitem [{\citenamefont {Jalas}\ \emph {et~al.}(2013)\citenamefont {Jalas},
  \citenamefont {Petrov}, \citenamefont {Eich}, \citenamefont {Freude},
  \citenamefont {Fan}, \citenamefont {Yu}, \citenamefont {Baets}, \citenamefont
  {Popovic}, \citenamefont {Melloni}, \citenamefont {Joannopoulos},
  \citenamefont {Vanwolleghem}, \citenamefont {Doerr},\ and\ \citenamefont
  {Renner}}]{JalaPetrEich13}%
  \BibitemOpen
  \bibfield  {author} {\bibinfo {author} {\bibfnamefont {D.}~\bibnamefont
  {Jalas}}, \bibinfo {author} {\bibfnamefont {A.}~\bibnamefont {Petrov}},
  \bibinfo {author} {\bibfnamefont {M.}~\bibnamefont {Eich}}, \bibinfo {author}
  {\bibfnamefont {W.}~\bibnamefont {Freude}}, \bibinfo {author} {\bibfnamefont
  {S.}~\bibnamefont {Fan}}, \bibinfo {author} {\bibfnamefont {Z.}~\bibnamefont
  {Yu}}, \bibinfo {author} {\bibfnamefont {R.}~\bibnamefont {Baets}}, \bibinfo
  {author} {\bibfnamefont {M.}~\bibnamefont {Popovic}}, \bibinfo {author}
  {\bibfnamefont {A.}~\bibnamefont {Melloni}}, \bibinfo {author} {\bibfnamefont
  {J.~D.}\ \bibnamefont {Joannopoulos}}, \bibinfo {author} {\bibfnamefont
  {M.}~\bibnamefont {Vanwolleghem}}, \bibinfo {author} {\bibfnamefont {C.~R.}\
  \bibnamefont {Doerr}}, \ and\ \bibinfo {author} {\bibfnamefont
  {H.}~\bibnamefont {Renner}},\ }\bibfield  {title} {\enquote {\bibinfo {title}
  {What is --- and what is not --- an optical isolator},}\ }\href
  {http://dx.doi.org/10.1038/nphoton.2013.185} {\bibfield  {journal} {\bibinfo
  {journal} {Nature Photonics}\ }\textbf {\bibinfo {volume} {7}},\ \bibinfo
  {pages} {579} (\bibinfo {year} {2013})}\BibitemShut {NoStop}%
\bibitem [{\citenamefont {Kosako}\ \emph {et~al.}(2010)\citenamefont {Kosako},
  \citenamefont {Kadoya},\ and\ \citenamefont {Hofmann}}]{Kosako:2010aa}%
  \BibitemOpen
  \bibfield  {author} {\bibinfo {author} {\bibfnamefont {T.}~\bibnamefont
  {Kosako}}, \bibinfo {author} {\bibfnamefont {Y.}~\bibnamefont {Kadoya}}, \
  and\ \bibinfo {author} {\bibfnamefont {H.~F.}\ \bibnamefont {Hofmann}},\
  }\bibfield  {title} {\enquote {\bibinfo {title} {Directional control of light
  by a nano-optical yagi-uda antenna},}\ }\href
  {http://dx.doi.org/10.1038/nphoton.2010.34} {\bibfield  {journal} {\bibinfo
  {journal} {Nature Photonics}\ }\textbf {\bibinfo {volume} {4}},\ \bibinfo
  {pages} {312} (\bibinfo {year} {2010})}\BibitemShut {NoStop}%
\bibitem [{\citenamefont {Zeeb}\ \emph {et~al.}(2015)\citenamefont {Zeeb},
  \citenamefont {Noh}, \citenamefont {Parkins},\ and\ \citenamefont
  {Carmichael}}]{Zeeb:PRA:2015}%
  \BibitemOpen
  \bibfield  {author} {\bibinfo {author} {\bibfnamefont {S.}~\bibnamefont
  {Zeeb}}, \bibinfo {author} {\bibfnamefont {C.}~\bibnamefont {Noh}}, \bibinfo
  {author} {\bibfnamefont {A.~S.}\ \bibnamefont {Parkins}}, \ and\ \bibinfo
  {author} {\bibfnamefont {H.~J.}\ \bibnamefont {Carmichael}},\ }\bibfield
  {title} {{\selectlanguage {English}\enquote {\bibinfo {title} {{Superradiant
  decay and dipole-dipole interaction of distant atoms in a two-way cascaded
  cavity QED system}},}\ }}\href {\doibase 10.1103/PhysRevA.91.023829}
  {\bibfield  {journal} {\bibinfo  {journal} {Physical Review A}\ }\textbf
  {\bibinfo {volume} {91}},\ \bibinfo {pages} {023829} (\bibinfo {year}
  {2015})}\BibitemShut {NoStop}%
\bibitem [{\citenamefont {Combes}\ \emph {et~al.}(2017)\citenamefont {Combes},
  \citenamefont {Kerckhoff},\ and\ \citenamefont {Sarovar}}]{CombKercSaro16}%
  \BibitemOpen
  \bibfield  {author} {\bibinfo {author} {\bibfnamefont {J.}~\bibnamefont
  {Combes}}, \bibinfo {author} {\bibfnamefont {J.}~\bibnamefont {Kerckhoff}}, \
  and\ \bibinfo {author} {\bibfnamefont {M.}~\bibnamefont {Sarovar}},\
  }\bibfield  {title} {\enquote {\bibinfo {title} {{The SLH framework for
  modeling quantum input-output networks}},}\ }\href {\doibase
  10.1080/23746149.2017.1343097} {\bibfield  {journal} {\bibinfo  {journal}
  {Advances in Physics: X}\ }\textbf {\bibinfo {volume} {2}},\ \bibinfo {pages}
  {784} (\bibinfo {year} {2017})}\BibitemShut {NoStop}%
\bibitem [{\citenamefont {Lalumi\`ere}\ \emph {et~al.}(2013)\citenamefont
  {Lalumi\`ere}, \citenamefont {Sanders}, \citenamefont {van Loo},
  \citenamefont {Fedorov}, \citenamefont {Wallraff},\ and\ \citenamefont
  {Blais}}]{LaluSandLoo13}%
  \BibitemOpen
  \bibfield  {author} {\bibinfo {author} {\bibfnamefont {K.}~\bibnamefont
  {Lalumi\`ere}}, \bibinfo {author} {\bibfnamefont {B.~C.}\ \bibnamefont
  {Sanders}}, \bibinfo {author} {\bibfnamefont {A.~F.}\ \bibnamefont {van
  Loo}}, \bibinfo {author} {\bibfnamefont {A.}~\bibnamefont {Fedorov}},
  \bibinfo {author} {\bibfnamefont {A.}~\bibnamefont {Wallraff}}, \ and\
  \bibinfo {author} {\bibfnamefont {A.}~\bibnamefont {Blais}},\ }\bibfield
  {title} {\enquote {\bibinfo {title} {Input-output theory for waveguide {QED}
  with an ensemble of inhomogeneous atoms},}\ }\href
  {https://dx.doi.org/10.1103/PhysRevA.88.043806} {\bibfield  {journal}
  {\bibinfo  {journal} {Phys. Rev. A}\ }\textbf {\bibinfo {volume} {88}},\
  \bibinfo {pages} {043806} (\bibinfo {year} {2013})}\BibitemShut {NoStop}%
\bibitem [{\citenamefont {Redchenko}\ and\ \citenamefont
  {Yudson}(2014)}]{RedcYuds14}%
  \BibitemOpen
  \bibfield  {author} {\bibinfo {author} {\bibfnamefont {E.~S.}\ \bibnamefont
  {Redchenko}}\ and\ \bibinfo {author} {\bibfnamefont {V.~I.}\ \bibnamefont
  {Yudson}},\ }\bibfield  {title} {\enquote {\bibinfo {title} {Decay of
  metastable states of two qubits in a waveguide},}\ }\href
  {https://dx.doi.org/10.1103/PhysRevA.90.063829} {\bibfield  {journal}
  {\bibinfo  {journal} {Physical Review A}\ }\textbf {\bibinfo {volume} {90}},\
  \bibinfo {pages} {063829} (\bibinfo {year} {2014})}\BibitemShut {NoStop}%
\bibitem [{\citenamefont {Nielsen}\ and\ \citenamefont
  {Chuang}(2000)}]{nielsen_2000}%
  \BibitemOpen
  \bibfield  {author} {\bibinfo {author} {\bibfnamefont {M.}~\bibnamefont
  {Nielsen}}\ and\ \bibinfo {author} {\bibfnamefont {I.}~\bibnamefont
  {Chuang}},\ }\href@noop {} {\emph {\bibinfo {title} {Quantum Computation and
  Quantum Information}}}\ (\bibinfo  {publisher} {Cambridge University Press},\
  \bibinfo {year} {2000})\BibitemShut {NoStop}%
\bibitem [{\citenamefont {Bouten}\ and\ \citenamefont
  {Silberfarb}(2008)}]{BoutSilb08}%
  \BibitemOpen
  \bibfield  {author} {\bibinfo {author} {\bibfnamefont {L.}~\bibnamefont
  {Bouten}}\ and\ \bibinfo {author} {\bibfnamefont {A.}~\bibnamefont
  {Silberfarb}},\ }\bibfield  {title} {\enquote {\bibinfo {title} {Adiabatic
  elimination in quantum stochastic models},}\ }\href
  {https://doi.org/10.1007/s00220-008-0513-6} {\bibfield  {journal} {\bibinfo
  {journal} {Communications in Mathematical Physics}\ }\textbf {\bibinfo
  {volume} {283}},\ \bibinfo {pages} {491} (\bibinfo {year}
  {2008})}\BibitemShut {NoStop}%
\bibitem [{\citenamefont {Bouten}\ \emph {et~al.}(2008)\citenamefont {Bouten},
  \citenamefont {van Handel},\ and\ \citenamefont
  {Silberfarb}}]{BoutHandSilb08}%
  \BibitemOpen
  \bibfield  {author} {\bibinfo {author} {\bibfnamefont {L.}~\bibnamefont
  {Bouten}}, \bibinfo {author} {\bibfnamefont {R.}~\bibnamefont {van Handel}},
  \ and\ \bibinfo {author} {\bibfnamefont {A.}~\bibnamefont {Silberfarb}},\
  }\bibfield  {title} {\enquote {\bibinfo {title} {Approximation and limit
  theorems for quantum stochastic models with unbounded coefficients},}\ }\href
  {http://dx.doi.org/10.1016/j.jfa.2008.02.013} {\bibfield  {journal} {\bibinfo
   {journal} {Journal of Functional Analysis}\ }\textbf {\bibinfo {volume}
  {254}},\ \bibinfo {pages} {3123} (\bibinfo {year} {2008})}\BibitemShut
  {NoStop}%
\bibitem [{\citenamefont {Stace}\ \emph {et~al.}(2005)\citenamefont {Stace},
  \citenamefont {Doherty},\ and\ \citenamefont
  {Barrett}}]{stace2005population}%
  \BibitemOpen
  \bibfield  {author} {\bibinfo {author} {\bibfnamefont {T.~M.}\ \bibnamefont
  {Stace}}, \bibinfo {author} {\bibfnamefont {A.~C.}\ \bibnamefont {Doherty}},
  \ and\ \bibinfo {author} {\bibfnamefont {S.~D.}\ \bibnamefont {Barrett}},\
  }\bibfield  {title} {\enquote {\bibinfo {title} {Population inversion of a
  driven two-level system in a structureless bath},}\ }\href {\doibase
  http://dx.doi.org/10.1103/PhysRevLett.95.106801} {\bibfield  {journal}
  {\bibinfo  {journal} {Physical Review Letters}\ }\textbf {\bibinfo {volume}
  {95}},\ \bibinfo {pages} {106801} (\bibinfo {year} {2005})}\BibitemShut
  {NoStop}%
\bibitem [{\citenamefont {Hughes}\ and\ \citenamefont
  {Carmichael}(2011)}]{Hughes:PRL:2011}%
  \BibitemOpen
  \bibfield  {author} {\bibinfo {author} {\bibfnamefont {S.}~\bibnamefont
  {Hughes}}\ and\ \bibinfo {author} {\bibfnamefont {H.~J.}\ \bibnamefont
  {Carmichael}},\ }\bibfield  {title} {{\selectlanguage {English}\enquote
  {\bibinfo {title} {{Stationary Inversion of a Two Level System Coupled to an
  Off-Resonant Cavity with Strong Dissipation}},}\ }}\href {\doibase
  10.1103/PhysRevLett.107.193601} {\bibfield  {journal} {\bibinfo  {journal}
  {Physical Review Letters}\ }\textbf {\bibinfo {volume} {107}},\ \bibinfo
  {pages} {193601} (\bibinfo {year} {2011})}\BibitemShut {NoStop}%
\bibitem [{\citenamefont {Hughes}\ and\ \citenamefont
  {Carmichael}(2013)}]{Hughes:NJP:2013}%
  \BibitemOpen
  \bibfield  {author} {\bibinfo {author} {\bibfnamefont {S.}~\bibnamefont
  {Hughes}}\ and\ \bibinfo {author} {\bibfnamefont {H.~J.}\ \bibnamefont
  {Carmichael}},\ }\bibfield  {title} {{\selectlanguage {English}\enquote
  {\bibinfo {title} {{Phonon-mediated population inversion in a semiconductor
  quantum-dot cavity system}},}\ }}\href {\doibase
  10.1088/1367-2630/15/5/053039} {\bibfield  {journal} {\bibinfo  {journal}
  {New Journal of Physics}\ }\textbf {\bibinfo {volume} {15}},\ \bibinfo
  {pages} {053039} (\bibinfo {year} {2013})}\BibitemShut {NoStop}%
\bibitem [{\citenamefont {Stace}\ \emph {et~al.}(2013)\citenamefont {Stace},
  \citenamefont {Doherty},\ and\ \citenamefont {Reilly}}]{stace2013dynamical}%
  \BibitemOpen
  \bibfield  {author} {\bibinfo {author} {\bibfnamefont {T.~M.}\ \bibnamefont
  {Stace}}, \bibinfo {author} {\bibfnamefont {A.~C.}\ \bibnamefont {Doherty}},
  \ and\ \bibinfo {author} {\bibfnamefont {D.~J.}\ \bibnamefont {Reilly}},\
  }\bibfield  {title} {\enquote {\bibinfo {title} {Dynamical steady states in
  driven quantum systems},}\ }\href {\doibase 10.1103/PhysRevLett.111.180602}
  {\bibfield  {journal} {\bibinfo  {journal} {Phys. Rev. Lett.}\ }\textbf
  {\bibinfo {volume} {111}},\ \bibinfo {pages} {180602} (\bibinfo {year}
  {2013})}\BibitemShut {NoStop}%
\bibitem [{\citenamefont {Colless}\ \emph {et~al.}(2014)\citenamefont
  {Colless}, \citenamefont {Croot}, \citenamefont {Stace}, \citenamefont
  {Doherty}, \citenamefont {Barrett}, \citenamefont {Lu}, \citenamefont
  {Gossard},\ and\ \citenamefont {Reilly}}]{Colless:2014aa}%
  \BibitemOpen
  \bibfield  {author} {\bibinfo {author} {\bibfnamefont {J.~I.}\ \bibnamefont
  {Colless}}, \bibinfo {author} {\bibfnamefont {X.~G.}\ \bibnamefont {Croot}},
  \bibinfo {author} {\bibfnamefont {T.~M.}\ \bibnamefont {Stace}}, \bibinfo
  {author} {\bibfnamefont {A.~C.}\ \bibnamefont {Doherty}}, \bibinfo {author}
  {\bibfnamefont {S.~D.}\ \bibnamefont {Barrett}}, \bibinfo {author}
  {\bibfnamefont {H.}~\bibnamefont {Lu}}, \bibinfo {author} {\bibfnamefont
  {A.~C.}\ \bibnamefont {Gossard}}, \ and\ \bibinfo {author} {\bibfnamefont
  {D.~J.}\ \bibnamefont {Reilly}},\ }\bibfield  {title} {\enquote {\bibinfo
  {title} {Raman phonon emission in a driven double quantum dot},}\ }\href
  {http://dx.doi.org/10.1038/ncomms4716} {\bibfield  {journal} {\bibinfo
  {journal} {Nature Communications}\ }\textbf {\bibinfo {volume} {5}},\
  \bibinfo {pages} {3716 EP } (\bibinfo {year} {2014})}\BibitemShut {NoStop}%
\bibitem [{\citenamefont {Peropadre}\ \emph {et~al.}(2013)\citenamefont
  {Peropadre}, \citenamefont {Lindkvist}, \citenamefont {Hoi}, \citenamefont
  {Wilson}, \citenamefont {Garcia-Ripoll}, \citenamefont {Delsing},\ and\
  \citenamefont {Johansson}}]{1367-2630-15-3-035009}%
  \BibitemOpen
  \bibfield  {author} {\bibinfo {author} {\bibfnamefont {B.}~\bibnamefont
  {Peropadre}}, \bibinfo {author} {\bibfnamefont {J.}~\bibnamefont
  {Lindkvist}}, \bibinfo {author} {\bibfnamefont {I.-C.}\ \bibnamefont {Hoi}},
  \bibinfo {author} {\bibfnamefont {C.~M.}\ \bibnamefont {Wilson}}, \bibinfo
  {author} {\bibfnamefont {J.~J.}\ \bibnamefont {Garcia-Ripoll}}, \bibinfo
  {author} {\bibfnamefont {P.}~\bibnamefont {Delsing}}, \ and\ \bibinfo
  {author} {\bibfnamefont {G.}~\bibnamefont {Johansson}},\ }\bibfield  {title}
  {\enquote {\bibinfo {title} {Scattering of coherent states on a single
  artificial atom},}\ }\href {http://stacks.iop.org/1367-2630/15/i=3/a=035009}
  {\bibfield  {journal} {\bibinfo  {journal} {New Journal of Physics}\ }\textbf
  {\bibinfo {volume} {15}},\ \bibinfo {pages} {035009} (\bibinfo {year}
  {2013})}\BibitemShut {NoStop}%
\bibitem [{\citenamefont {Lu}\ \emph {et~al.}(2014)\citenamefont {Lu},
  \citenamefont {Zhou}, \citenamefont {Kuang},\ and\ \citenamefont
  {Nori}}]{PhysRevA.89.013805}%
  \BibitemOpen
  \bibfield  {author} {\bibinfo {author} {\bibfnamefont {J.}~\bibnamefont
  {Lu}}, \bibinfo {author} {\bibfnamefont {L.}~\bibnamefont {Zhou}}, \bibinfo
  {author} {\bibfnamefont {L.-M.}\ \bibnamefont {Kuang}}, \ and\ \bibinfo
  {author} {\bibfnamefont {F.}~\bibnamefont {Nori}},\ }\bibfield  {title}
  {\enquote {\bibinfo {title} {Single-photon router: Coherent control of
  multichannel scattering for single photons with quantum interferences},}\
  }\href {\doibase 10.1103/PhysRevA.89.013805} {\bibfield  {journal} {\bibinfo
  {journal} {Phys. Rev. A}\ }\textbf {\bibinfo {volume} {89}},\ \bibinfo
  {pages} {013805} (\bibinfo {year} {2014})}\BibitemShut {NoStop}%
\bibitem [{Note1()}]{Note1}%
  \BibitemOpen
  \bibinfo {note} {Such a scheme is physically realisable by using a {QND}
  detection of photon flux to determine from which direction the field is
  incident, and controlling the mirror accordingly}\BibitemShut {NoStop}%
\bibitem [{\citenamefont {Stannigel}\ \emph {et~al.}(2012)\citenamefont
  {Stannigel}, \citenamefont {Rabl},\ and\ \citenamefont
  {Zoller}}]{StanRablZoll12}%
  \BibitemOpen
  \bibfield  {author} {\bibinfo {author} {\bibfnamefont {K.}~\bibnamefont
  {Stannigel}}, \bibinfo {author} {\bibfnamefont {P.}~\bibnamefont {Rabl}}, \
  and\ \bibinfo {author} {\bibfnamefont {P.}~\bibnamefont {Zoller}},\
  }\bibfield  {title} {\enquote {\bibinfo {title} {Driven-dissipative
  preparation of entangled states in cascaded quantum-optical networks},}\
  }\href {http://dx.doi.org/10.1088/1367-2630/14/6/063014} {\bibfield
  {journal} {\bibinfo  {journal} {New Journal of Physics}\ }\textbf {\bibinfo
  {volume} {14}},\ \bibinfo {pages} {063014} (\bibinfo {year}
  {2012})}\BibitemShut {NoStop}%
\bibitem [{\citenamefont {Ramos}\ \emph {et~al.}(2014)\citenamefont {Ramos},
  \citenamefont {Pichler}, \citenamefont {Daley},\ and\ \citenamefont
  {Zoller}}]{RamoPichDale14}%
  \BibitemOpen
  \bibfield  {author} {\bibinfo {author} {\bibfnamefont {T.}~\bibnamefont
  {Ramos}}, \bibinfo {author} {\bibfnamefont {H.}~\bibnamefont {Pichler}},
  \bibinfo {author} {\bibfnamefont {A.~J.}\ \bibnamefont {Daley}}, \ and\
  \bibinfo {author} {\bibfnamefont {P.}~\bibnamefont {Zoller}},\ }\bibfield
  {title} {\enquote {\bibinfo {title} {Quantum spin dimers from chiral
  dissipation in cold-atom chains},}\ }\href
  {https://dx.doi.org/10.1103/PhysRevLett.113.237203} {\bibfield  {journal}
  {\bibinfo  {journal} {Phys. Rev. Lett.}\ }\textbf {\bibinfo {volume} {113}},\
  \bibinfo {pages} {237203} (\bibinfo {year} {2014})}\BibitemShut {NoStop}%
\bibitem [{\citenamefont {Pichler}\ \emph {et~al.}(2015)\citenamefont
  {Pichler}, \citenamefont {Ramos}, \citenamefont {Daley},\ and\ \citenamefont
  {Zoller}}]{PichRamoDale15}%
  \BibitemOpen
  \bibfield  {author} {\bibinfo {author} {\bibfnamefont {H.}~\bibnamefont
  {Pichler}}, \bibinfo {author} {\bibfnamefont {T.}~\bibnamefont {Ramos}},
  \bibinfo {author} {\bibfnamefont {A.~J.}\ \bibnamefont {Daley}}, \ and\
  \bibinfo {author} {\bibfnamefont {P.}~\bibnamefont {Zoller}},\ }\bibfield
  {title} {\enquote {\bibinfo {title} {Quantum optics of chiral spin
  networks},}\ }\href {https://dx.doi.org/10.1103/PhysRevA.91.042116}
  {\bibfield  {journal} {\bibinfo  {journal} {Phys. Rev. A}\ }\textbf {\bibinfo
  {volume} {91}},\ \bibinfo {pages} {042116} (\bibinfo {year}
  {2015})}\BibitemShut {NoStop}%
\bibitem [{\citenamefont {Gardiner}\ and\ \citenamefont
  {Zoller}(2000)}]{GardZoll00}%
  \BibitemOpen
  \bibfield  {author} {\bibinfo {author} {\bibfnamefont {C.~W.}\ \bibnamefont
  {Gardiner}}\ and\ \bibinfo {author} {\bibfnamefont {P.}~\bibnamefont
  {Zoller}},\ }\href {https://www.springer.com/gp/book/9783540223016} {\emph
  {\bibinfo {title} {Quantum Noise}}},\ \bibinfo {edition} {2nd}\ ed.\
  (\bibinfo  {publisher} {Springer},\ \bibinfo {year} {2000})\BibitemShut
  {NoStop}%
\end{thebibliography}%

\clearpage

\appendix



\section{SLH modeling}\label{app:slh}
We start from the SLH triple in Eq.~\eqref{eqn:lk} for each atomic systems, which are cascaded with two coherent state source models \cite{CombKercSaro16}, leading to Equations \eqref{eq:HSLH}--\eqref{eq:L2SLH} in the main text.  
(See also Eq.\ (132) and Eqs.\ (174) - (176) of \cite{CombKercSaro16}, for additional discussion). 
 
We may re-factor the SLH master equation to obtain a more accessible form of the operators
\begin{align}
	\dot{\rho} =& \mathcal L \rho = -\ii \comm{H}{\rho} + \diss{\bar L_1}\rho + \diss{\bar L_2}\rho \,, \\
	H =& H_0 + H_C + H_D \,,
\end{align}
with the operators
\begin{align}
	H_0 =& H_1 + H_2 \,, \\
	H_C =& \frac12 \sqrt{\gamma_1 \gamma_2} \sin{\phi} \l( \sigma_-^{(1)} \sigma_+^{(2)} + \sigma_+^{(1)} \sigma_-^{(2)} \r) \,,\\
	H_D =& \ii \sqrt{\frac{\gamma_1}{2}} \big( (\alpha\cc + \ee^{-\ii \phi} \beta\cc ) \sigma_-^{(1)} - (\alpha + \ee^{\ii \phi} \beta) \sigma_+^{(1)} \big) \nn\\
		+& \ii \sqrt{\frac{\gamma_2}{2}} \big((\beta\cc + \ee^{-\ii \phi} \alpha\cc ) \sigma_-^{(2)} - (\beta + \ee^{\ii \phi} \alpha) \sigma_+^{(2)} \big)\,,\\
	\bar L_1 =& L_1 + \ee^{\ii \phi} L_2 \,,\\
	\bar L_2 =& L_2 + \ee^{\ii \phi} L_1 \,,
\end{align}
where $H_{C}$ describes coupling between the two atoms, mediated by the field and $H_{D}$ is the coherent drive acting on each atom. The modified Lindblad operators $\bar L_{1/2}$ now describe purely the decay of the atoms into the field. 
Note that in order to correctly calculate output field amplitudes and fluxes, we still need to keep in mind the original Lindblad operators, Eqs.~\eqref{eq:L1SLH}-\eqref{eq:L2SLH}. 
However, the above description correctly reproduces the dynamics of the atomic degrees of freedom, i.e. the master equations is equivalent to Eq.~\eqref{eq:ME0}.

\section{Hamiltonian in dark / bright state basis}\label{app:symmbasis}

Defining the bright and dark states as the symmetric and antisymmetric superposition of the states with a single atomic excitation
\begin{align}
	\ket{B} &= \frac{1}{\sqrt2} \l( \ket{ge} - \ket{eg} \r) \quad,\quad \ket{D}= \frac{1}{\sqrt2} \l(  \ket{ge} + \ket{eg} \r) \,,
\end{align}
we can define new ladder operators as
\begin{align}
	\sigma_{-}^{(B)} &= \frac{1}{\sqrt2} \l( \sigma_{-}^{(2)} -  \sigma_{-}^{(1)} \r) \quad, \quad \sigma_{-}^{(D)} = \frac{1}{\sqrt2} \l( \sigma_{-}^{(2)} +  \sigma_{-}^{(1)} \r) \,,
\end{align}
with $\sigma_{+}^{(B)}\ket{gg} = \ket B$ and $\sigma_{+}^{(D)} \ket{gg} = \ket D$.
In the basis $\l\{ \ket{gg}, \ket{D}, \ket{B}, \ket{ee} \r\}$, we can then write the Hamiltonian of the atoms and their effective coupling as 
\begin{align}
	H_0 + H_C &\approx 
	\frac12 \l(	\begin{array}{cccc}
		-\omega_{1} - \omega_{2} & 0 & 0& 0\\
		0 & -g_{0}\sin{\phi} & \omega_{2}-\omega_{1} &0 \\
		0 & \omega_{2}-\omega_{1} & g_{0}\sin\phi & 0\\
		0 & 0& 0& \omega_{1} + \omega_{2} 
	\end{array} \r) \,,
\end{align}
where $g_{0} = \sqrt{\gamma_{1}\gamma_{2}}$. 
This expression makes evident that in the dark / bright state basis, a detuning between the two atoms, $\delta\omega = \omega_{1} - \omega_{2} \neq 0$, leads to an effective coupling between the symmetric and the antisymmetric state.
Conversely, a coupling term between the atoms will lead to a splitting between the dark and bright state.
Assuming equal coupling of the two atoms to the waveguide, $\gamma_{1} = \gamma_{2} = \gamma$, we find for the driving terms and dissipative operators
\begin{align}
	H_{D} &= -\ii \frac{\sqrt{\gamma}}{2} \Bigl\{ \sigma_{+}^{(D)} (\alpha+\beta)(1+\ee^{\ii\phi}) \nn\\
		&\quad\quad\quad\quad - \sigma_{+}^{(B)} (\alpha-\beta)(1-\ee^{\ii\phi}) \Bigr\} + \text{h.c.} ,\\
	\bar L_{1} &= \frac{\sqrt\gamma}{2} \Bigl\{ \sigma_{-}^{(D)} (1+\ee^{\ii\phi}) - \sigma_{-}^{(B)} (1-\ee^{\ii \phi}) \Bigr\} \,,\nn\\
	\bar L_{2} &= \frac{\sqrt\gamma}{2} \Bigl\{ \sigma_{-}^{(D)} (1+\ee^{\ii\phi}) + \sigma_{-}^{(B)} (1-\ee^{\ii \phi})\Bigr\} \,.
\end{align}

As discussed bellow and in the main text, 
in the following we adopt optimal parameters \mbox{$\omega_{1} = -\delta\,\gamma$}, $\omega_{2} = 0$ and $\phi = \pi - \delta$, with $\delta \ll \gamma=1$, and we will only write the results to lowest non-trivial order in $\delta$.

Then we find for the Hamiltonian of the atoms plus their coupling
\begin{align}
	H_0 + H_C &\approx 
	\frac12 \gamma \l(	\begin{array}{cccc}
		\delta & 0 & 0& 0\\
		0 & \delta & \delta &0 \\
		0 & \delta & -\delta & 0\\
		0 & 0& 0& -\delta 
	\end{array} \r) .
\end{align}
Defining the small parameter $\epsilon = \frac12 \ii \delta $, we can write the driving Hamiltonian as
\begin{align}
	H_{D} \approx &\phantom{-} \ii \sqrt{\gamma} \alpha \Bigl\{ \l( 1- \epsilon \r) \sigma_{+}^{(B)} - \epsilon \sigma_{+}^{(D)} \Bigr\} \nn\\
		& -\ii \sqrt{\gamma} \beta \Bigl\{  \l( 1-\epsilon \r) \sigma_{+}^{(B)} + \epsilon \sigma_{+}^{(D)}   \Bigr\} + \text{h.c.} ,
\end{align}
while for the sum of the two dissipators we find 
\begin{align}
	\diss{\bar L_{1}}\rho &+ \diss{\bar L_{2}} \rho \nn\\
	&= 2\gamma\abss{1 - \epsilon} \diss{\sigma_{-}^{(B)}}\rho + 2\gamma \abss \epsilon \diss{\sigma_{-}^{(D)}}\rho \,.
\end{align}
To second order in $\delta$ we thus find the decay rates of the bright and dark states
\begin{align}
	\gamma_{B} &= 2 \abss{1 - \epsilon}\gamma \approx  2 \gamma \,, \nn\\
	\gamma_{D} &= 2 \abss \epsilon \gamma \approx {\delta^{2}\gamma}/{2}.
\end{align}
Thus the antisymmetric state has a fast decay rate, so that can be identified as the bright state, and the symmetric state has a slow decay rate, making it the dark state.

\section{Optimised efficiency}\label{app:optimum}

\begin{figure}[t]
	\begin{center}
	\includegraphics[width=\columnwidth]{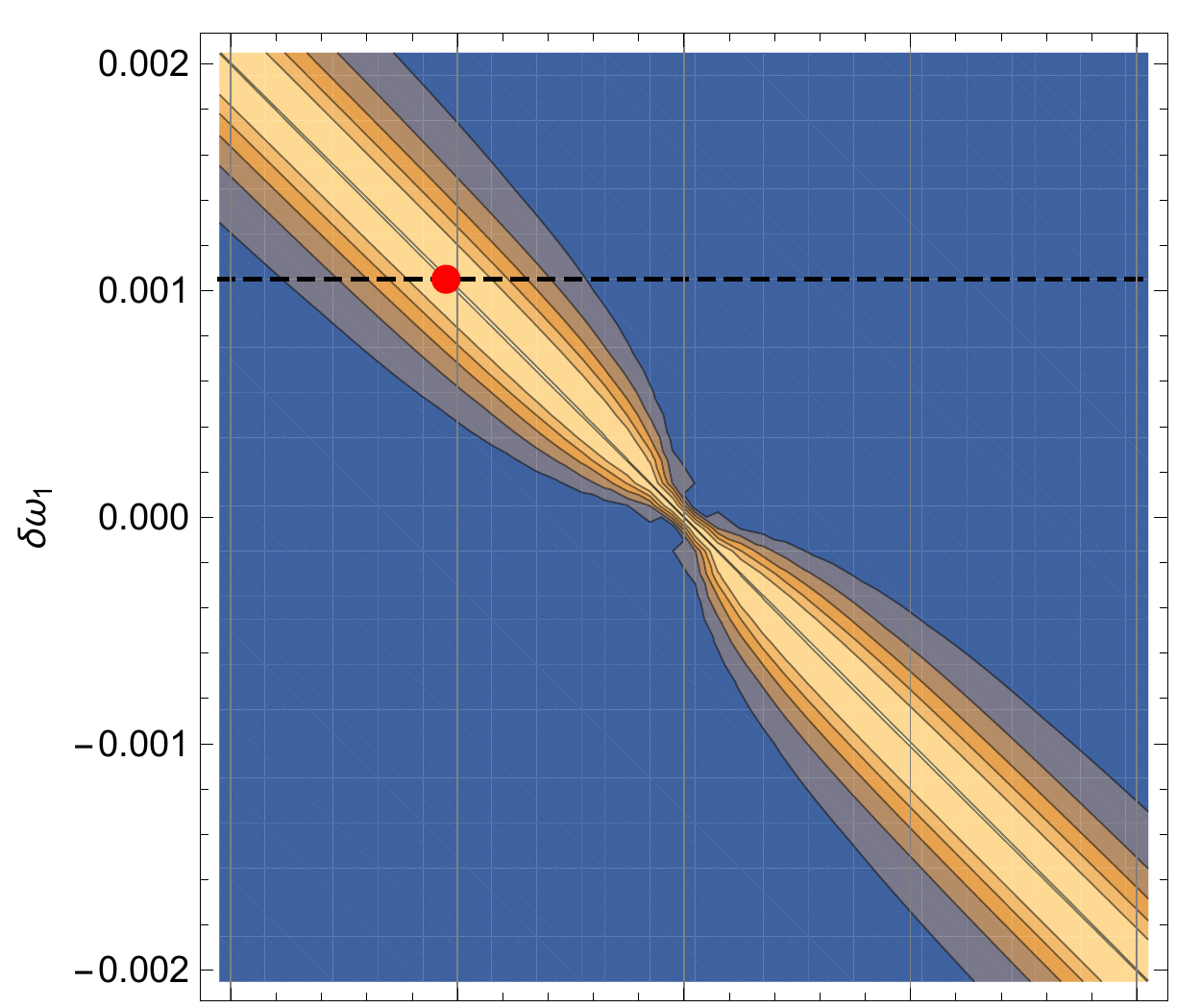}
	\includegraphics[width=\columnwidth]{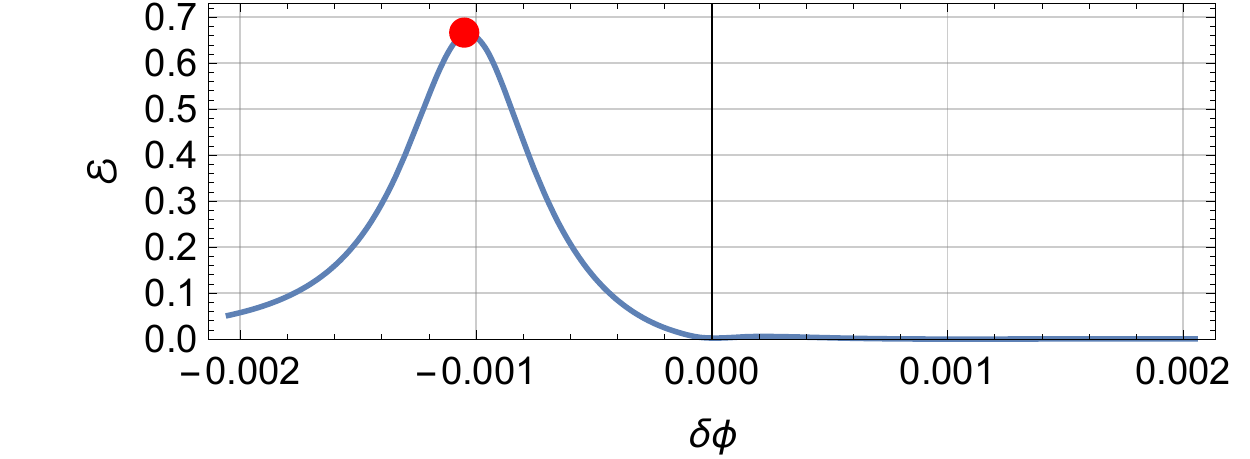}
		\caption{(colour online) (Top) Contour plot of the `diode efficiency', $\mathcal{E}$ as a function of inter-atom path length $\delta\phi$, and the detuning of one atom $\delta\omega_1$, assuming $\delta\omega_2=0$.  
		(Bottom) Line cut through contour plot along dashed line.  The peak efficiency, $\mathcal{E}=2/3$ occurs along the line $\delta\phi=-\delta\omega_1\equiv\delta$.  
		Allowing $\delta\omega_2$ to vary does not yield a higher value for $\mathcal{E}$.
	}
	\label{fig:effplots}
	\end{center}
\end{figure}

Throughout the preceding analysis, we considered \mbox{$\delta\phi=-\delta\omega_1\equiv\delta$}, and $\delta\omega_2=0$.
This choice of parameters optimises the `diode efficiency' \mbox{$\mathcal{E}=\mathcal{T}_\alpha(\mathcal{T}_\alpha-\mathcal{T}_\beta)/(\mathcal{T}_\alpha+\mathcal{T}_\beta)$}, 
which is a measure of the left-right asymmetry in the flux transmission.  
To demonstrate this, we numerically tabulate $\mathcal{E}$ over atomic detunings, $\delta\omega_1$ and $\delta\omega_2$ for each value of $\delta\phi$ .  
We find the optimal left-right asymmetry when $\delta\omega_2=0$, and \mbox{$\delta\phi=-\delta\omega_1\equiv\delta$}.  
Part of this numerical calculation is shown in \fig{fig:effplots} in which we fix $\delta\omega_2=0$, and find that $\mathcal{E}_{\text{max}}\approx2/3$ along the optimal parameter choices above.  
The following analytical calculations, which adopt this parameter choice, confirm this as the maximum diode efficiency.

\section{Analytics along optimal parameters}\label{app:ss}


We solve for the steady-state of the master equation, $\mathcal L \rho_{ss} =0$, perturbatively  in $\delta$, using the expansion 
\begin{align}
	 \rho_{ss} &= \bar\rho_{0} + \ii \delta \bar \rho_{1} + \delta^{2} \bar \rho_{2} +\ldots\nonumber\\
		\mathcal L& = \mathcal L_{0} + \ii \delta \mathcal L_{1} + \delta^{2} \mathcal L_{2} + \ldots ,
\end{align}
requiring that the steady-state equation, $\mathcal L  \rho_{ss} =0$ is fullfilled at all orders of $\delta$ independently. 
This then leads to a hierarchy of equations for the components of the steady-state density matrix
\begin{align}
	0 &= \mathcal L_{0} \bar\rho_{0} \nn\\
	0 &= \mathcal L_{0} \bar\rho_{1} + \mathcal L_{1} \bar \rho_{0} \nn\\
	0 &= \mathcal L_{0} \bar\rho_{2} - \mathcal L_{1} \bar \rho_{1} + \mathcal L_{2} \bar\rho_{0} \nn\\
	&\ldots
	\label{eq:MEHierarchy}
\end{align}

Technically we always solve for the nullspace of a linear system of equations. 
We find that the nullspace of $\mathcal L_{0}$ is two-fold degenerate, so that we can write
\begin{align}
	\bar \rho_{0} = a_{1} \bar \rho_{0}^{(1)} + a_{2} \bar \rho_{0}^{(2)} \,, 
\end{align}
where we leave the coefficients $a_{i}$ arbitrary for the moment. We could fix one of the two coefficients by e.g. requiring the unit trace condition for a valid physical density matrix, but we will only do that later for clarity. The second parameter remains always undetermined at lowest oder $\delta$.

Plugging the zero-th order solution into the first order equation above, Eq.~\eqref{eq:MEHierarchy}, we find a four-fold degenerate nullspace of the first order equations.
Two of the solutions correspond to the choice $a_{1} = a_{2} = 0$ for $\bar\rho_{0}$, in which case the first order equation reduces to finding the nullspace of $\mathcal L_{0}$ again. 

The other two solutions are non-trivial and together this leads again to a parametrization of the first-order steady-state as
\begin{align}
	\bar \rho_{1} = b_{1} \bar \rho_{1}^{(1)} + b_{2} \bar \rho_{1}^{(2)} + b_{3} \bar \rho_{0}^{(1)} + b_{4} \bar \rho_{0}^{(2)} \,, 
\end{align}
where the coefficients are initially completely free. 
Fixing the trace at first order as $\text{Tr}\bar\rho_{1} = 0$ (to guarantee a valid density matrix for all $\delta$), allows us to fix one of the four parameters in $\bar\rho_{1}$. 
Two of the remaining three parameters can be determined by requiring that the above ansatz actually is a steady-state solution of the master equation to first order in $\delta$, i.e.
\begin{align}
	\l( \mathcal L_{0} + \ii \delta \mathcal L_{1} \r) \l( \bar\rho_{0} + \ii \delta\bar \rho_{1} \r) =0
\end{align}
to first order in $\delta$. 
At this point we are thus left with the free parameters $a_{1}, a_{2}$ and one of the $b_{i}$, where one of the $a_{i}$ can be determined from the unit trace condition.

Repeating the procedure for the second order equation in the above hierarchy, Eq.~\eqref{eq:MEHierarchy}, then finally allows us to uniquely determine the $a_{i}$. 
We find 
\begin{align}
	a_{2}^{(\alpha)} = \frac{1}{2\alpha^{4}} \l( 2+ \alpha^{2} + 2\alpha^{4} \r) a_{1}  \quad ,\quad a_{2}^{(\beta)} = \l( 1+\frac{1}{2\beta^{2}} \r) a_{1} \,,
\end{align}
with $a_{1}$ fixed by the trace.

Going to third order in the above hierarchy then uniquely fixes all parameters of the first order parametrisation, and we can write
\begin{align}
	\rho_{ss} \approx \bar\rho_{0} + \ii \delta\bar \rho_{1} 
\end{align}
which is different for driving from either side and coincides well with the numerical solutions for these parameters. 
For driving from the left, $\beta=0$, we have
\begin{widetext}
\begin{align}
	\bar\rho_{0}^{(\alpha)} &= \frac{1}{a_{0}} \l( \begin{array}{cccc} 
			2+2\alpha^{2} + 2\alpha^{4} & \sqrt{2}\alpha (1+\alpha^{2}) & -\sqrt{2} \alpha (1+\alpha^{2}) & -2\alpha^{2} \\
			\sqrt{2}\alpha (1+\alpha^{2}) & 2 + \alpha^{2}+ 2\alpha^{4} &2-\alpha^{2} & -\sqrt{2} \alpha^{3} \\
			-\sqrt{2}\alpha (1+\alpha^{2}) & 2-\alpha^{2} & 2+ \alpha^{2} + 2\alpha^{4} & \sqrt{2} \alpha^{3} \\
			-2\alpha^{2} & -\sqrt{2} \alpha^{3} & \sqrt{2} \alpha^{3} & 2\alpha^{4}\\
		 \end{array} \r) \,,\nn\\
	\bar\rho_{1}^{(\alpha)} &= \frac{1}{a_{0}} \l( \begin{array}{cccc}
			0 & \frac{\sqrt{2}}{\alpha} (1+\alpha^{4}) & \frac{\sqrt{2}}{\alpha} (1+\alpha^{2}+2\alpha^{4}) & 2\alpha^{2} \\
			-\frac{\sqrt{2}}{\alpha} (1+\alpha^{4}) & 0 & \alpha^{2} & \sqrt2 \alpha (1+2\alpha^{2}) \\
			-\frac{\sqrt{2}}{\alpha} (1+\alpha^{4}) & -\alpha^{2} & 0 & \sqrt2 \alpha (1+2\alpha^{2}) \\
			-2\alpha^{2} & -\sqrt2 \alpha (1+2\alpha^{2}) & - \sqrt2 \alpha (1+2\alpha^{2}) & 0\\
		\end{array} \r) \,, 
\end{align}

with $a_{0} = 6+4\alpha^{2} + 8\alpha^{4} $. 
For driving from the right ($\alpha=0$), we find
\begin{align}
	\bar\rho_{0}^{(\beta)} &= \frac{1}{b_{0}} \l( \begin{array}{cccc} 
			2(1+\beta^{2} + \beta^{4}) & -\sqrt2 \beta(1 + \beta^{2}) & \sqrt 2 \beta (1+\beta^{2}) & -2\beta^{2} \\
			-\sqrt2 \beta(1+\beta^{2}) &\beta^{2}+2\beta^{4} & -\beta^{2} & \sqrt2 \beta^{3} \\
			\sqrt2 \beta(1+\beta^{2}) &-\beta^{2} & \beta^{2} + 2\beta^{4} & -\sqrt2 \beta^{3} \\
			-2\beta^{2} & \sqrt2 \beta^{3} & -\sqrt2 \beta^{3} & 2 \beta^{4} 
		 \end{array} \r) \,,\nn\\
	\bar\rho_{1}^{(\beta)} &= \frac{1}{b_{0}} \l( \begin{array}{cccc}
			0 & 0 & -\sqrt 2 \beta(1+\beta^{2}) & 2\beta^{2} \\
			0 & 0 & \beta^{2} & -\sqrt2 \beta^{3} \\
			\sqrt 2 \beta(1+\beta^{2}) & -2\beta^{2} & 0 & 0\\
			-\beta^{2} & \sqrt2 \beta^{3} & 0 & 0 
		\end{array} \r) \,,
\end{align}
with $b_{0} = 2 \l( 1+2\beta^{2} + 4 \beta^{4} \r)$. 
\end{widetext}

It is instructive to consider the limit of the above expression for weak driving. To lowest order in $\alpha,\beta$, we find
\begin{align}
	\bar\rho^{(\alpha)} \approx & \frac{1}{9\sqrt2\alpha} \l( \begin{array}{cccc}
			1 & a_{1} & a_{2} & 0 \\
			a_{1}\cc & 1 & 1 & \ii 9\alpha^{2}\delta \\
			a_{2}\cc & 1 & 1 &  \ii 9\alpha^{2}\delta \\
			0 & -\ii 9 \alpha^{2}\delta & - \ii 9 \alpha^{2}\delta & 0\\
		\end{array} \r) \,,\nn\\
	\bar\rho^{\beta} \approx & \l( \begin{array}{cccc}
			1 & -\frac{\beta}{\sqrt2} & \frac{\beta}{\sqrt2} (1-\ii \delta) & 0 \\
			 -\frac{\beta}{\sqrt2} & 0 & 0 & 0\\
			  \frac{\beta}{\sqrt2} (1+\ii \delta) & 0 & 0 & 0 \\
			  0 & 0 & 0 & 0 \\
		\end{array} \r) \,,
\end{align}
with $a_{1}=  \alpha^{2}(3-2\ii\delta) + 3\ii \delta  $, $a_{2}= - \alpha^{2} (3 - \ii \delta) + 3\ii \delta $. 
Here the steady-state for driving from the left can be approximately expressed as $\bar\rho^{(\alpha)} \approx \frac13 \ket{00}\bra{00} + \frac23 \ket+\bra+$ and for driving from the right we find 
$\bar\rho^{(\beta)} \approx \ket{00}\bra{00}$ as stated in the main text.
These solutions are strictly valid only for $\delta < \alpha,\beta$.

\section{Adiabatic elimination in SLH formalism \label{app:AdiabElim}}

Adapting the treatment in Ref.~\cite{CombKercSaro16} (which elaborates on Refs.~\cite{BoutSilb08,BoutHandSilb08}), we perform adiabatic elimination directly for the SLH operator triplet.
To this end we define a \emph{slow} and \emph{fast} subspace via the projectors
\begin{align}
	\Pi_{0} = {\ket{G}}{\bra{G}} + \ket{D}\bra D \quad,\quad \Pi_{1} = \ket B \bra B + {\ket{E}}{\bra{E}}
\end{align}
where our choice is motivated through the observation that the steady state is limited to groundstate ${\ket{G}}$ and dark state $\ket D$.

Using the operator 
\begin{align}
	K = -\l( \ii H +\frac12 \sum_{k} L_{k}\hc L_{k} \r) = Y + A + B
\end{align}
we then define the fast, slow and intermediate parts of K through
\begin{align}
	Y &= \Pi_{1} K \Pi_{1} \,,\nn\\
	A &= \Pi_{1} K \Pi_{0} + \Pi_{0} K \Pi_{1} \,,\nn\\
	B &= \Pi_{0} K \Pi_{0} \,.
\end{align}
Performing a similar distinction for the Lindblad terms in the SLH triple
\begin{align}
	L_{k} = F_{k} + G_{k} \,,
\end{align}
with
\begin{align}
	F_{k} = \Pi_{1} L_{k} \Pi_{1} + \Pi_{0} L_{k} \Pi_{1} \,,\nn\\
	G_{k} = \Pi_{1} L_{k} \Pi_{0} + \Pi_{0} L_{k} \Pi_{0}\,.
\end{align}
we then get the equations for the SLH quantities in the adiabatically eliminated subspace defined by $\Pi_{0}$ as
\begin{align}
	\tilde K &= \Pi_{0} \l( B - A \tilde Y A \r) \Pi_{0} = -\l( \ii \tilde H + \frac12 \sum_{k} \tilde L_{k}\hc \tilde L_{k} \r) \,,\nn\\
	\tilde L_{k} &= \l( G_{k} - F_{k} \tilde Y A \r) \Pi_{0}
\end{align}
which allows us to extract the adiabatically eliminated Hamiltonian $\tilde H$ and Lindbladians $\tilde L_{k}$, as given in the main text.
Here the operator $\tilde Y$ is defined as the (pseudo)-inverse of Y with respect to the fast subspace, through
\begin{align}
	\tilde Y Y = Y \tilde Y = \Pi_{1} \,.
\end{align}
Also, since the scattering matrix S in our case is not relevant for any physical quantities of interest (since we already cascaded the source term into our original SLH triple), we also do not explicitely calculate the adiabatically eliminated $\tilde S$.

The projection method above is chosen such as to automatically satisfy the operator conditions that are necessary for the validity of the elimination scheme, namely
\begin{align}
	Y \Pi_{0} &= 0 \,,\nn\\
	F_{k} \Pi_{0} &= 0 \quad \forall k \,,\nn\\ 
	\Pi_{0} A \Pi_{0} &= 0\,.
\end{align}

\section{Field correlation functions\label{app:CorrFunc}}

Following Ref.~\cite{GardZoll00} we can calculate the output field correlation functions as
\begin{align}
	g^{(1)}(t_{0}, t_{1}) &= \frac{\mean{a\hc(t_{0}) a(t_{1})} }{ \sqrt{\mean{a\hc(t_{0}) a(t_{0})} \mean{a\hc(t_{1}) a(t_{1})}} } \,,\nn\\
	g^{(2)}(t_{0}, t_{1}) &= \frac{\mean{a\hc(t_{0}) a\hc(t_{1)} a(t_{1}) a(t_{0})} }{ \mean{a\hc(t_{0}) a(t_{0})} \mean{a\hc(t_{1}) a(t_{1})} } \,,\nn\\
\end{align}
where we take the initial time $t_{0}$ to be a time at which the system is already in equilibrium, i.e. has reached steady-state. 
Then the equal-time correlation functions are simply the equilibrium fluxes
\begin{align}
	&\mean{a\hc(t_{0}) a(t_{0})} = \mean{a\hc(t_{1}) a(t_{1})} \nn\\
		&\quad= \Tr \l\{ a\hc a \rho_{ss} \r\} = \Tr \l\{ a\hc a \ee^{\mathcal L t_{1}} \rho_{ss} \r\} \nn\\
		&\quad= \Tr \l\{ L\hc L \rho_{ss} \r\} 
\end{align}
with the steady-state density matrix $ \rho_{ss}$, the Liouvillian superoperator describing the dissipative time-evolution $\mathcal L$, and where we used the fact that $ \rho_{ss}$ is stationary under the action of $\mathcal L$.
Note that we replaced field operators by system/atom operators in the last line, in the usual input-output logic
\begin{align}
	a_{\text{out}} = a_{\text{in}} + \sqrt{\gamma} \sigma_{-} = L \,.
\end{align}
Two-time correlation functions we calculate according to~\cite{GardZoll00}
\begin{align}
	\mean{A(t_{0}) B(t_{1})} &= \Tr\l\{ B \ee^{\mathcal L(t_{1}-t_{0})} \rho(t_{0}) A \r\} \,,\nn\\
	\mean{A(t_{0}) B(t_{1}) C(t_{1}) D(t_{0})} &= \Tr \l\{  B C \ee^{\mathcal L (t_{1} - t_{0})} D \rho(t_{0}) A \r\} \,,
\end{align}
which translates into
\begin{align}
	\mean{a\hc(t_{0}) a(t_{1})} &= \Tr\l\{ L \ee^{\mathcal L t_{1}} \rho_{ss} L\hc \r\} \,,\nn\\
	\mean{a\hc(t_{0}) a\hc(t_{1}) a(t_{1}) a(t_{0)}} &= \Tr\l\{ L\hc L e^{\mathcal L t_{1}} L \rho_{ss} L\hc \r\} \,.
\end{align}



\section{Adiabatic elimination to obtain population rate equations \label{app:rate}}

As a prelude to the flapping mirror model, we derive a rate equation model for the dark and ground state populations, by adiabatically eliminating the off-diagonal elements $\rho_{DG}$ and $\rho_{GD}$ in \eqn{eqn:lk}, assuming the reduced Hamiltonian and Lindblad operators given in \eqn{eq:SLHElim}. 
In practice, we set $\dot \rho_{DG}=\dot \rho_{GD}=0$, and solve the resulting algebraic equations for  $\rho_{DG}$ and $\rho_{GD}$, in terms of the populations $P_G=\rho_{GG}$ and $P_D=\rho_{DD}$.  
Considering $\alpha$-driving (i.e.\ setting $\beta=0$), we find
\begin{align}
	\rho_{DG}=\rho_{GD}^*=\frac{-2i\,\alpha\,\delta}{2\alpha^2+\delta^2}\rho_{GG},
\end{align}
from which it follows that 
\begin{align}
	\l[ \begin{array}{c}
		\dot P_D \\
		\dot P_G
	\end{array} \r] &= \left[
	\begin{array}{cc}
		 -\delta^2 &  \frac{4\alpha^2\delta^2}{2\alpha^2+\delta^2}\\
		 \hphantom{-}\delta^2 & \frac{-4\alpha^2\delta^2}{2\alpha^2+\delta^2} \\
	\end{array}
	\right] \l[ \begin{array}{c}
		P_D \\
		P_G
	\end{array} \r],\\
	&= \left[
	\begin{array}{cc}
		 -\delta^2 &  \hphantom{-}2 \delta^2\\
		 \hphantom{-}\delta^2 & -2 \delta^2 \\
	\end{array}
	\right] \l[ \begin{array}{c}
		P_D \\
		P_G
	\end{array} \r]+O(\alpha^2)\label{eqn:dgrate}
\end{align}

\section{Flapping mirror model\label{app:Flap}}

The rate equations in \eqn{eqn:dgrate} correspond to a poisson process in which the two-atom system is fluctuates between states $\ket{G}$ and $\ket{D}$, with transition  rates given by the elements of the matrix in  \eqn{eqn:dgrate}.  
For low driving powers, the state $\ket{G}$ is reflective, while the state $\ket{D}$ is decoupled from the field so is transparent.

Thus we consider a toy model of a field propagating through a  flapping mirror which either reflects the input field $a_{\text{in}}$ to $a_{\text{out}}$, if it is in state $R=1$, or transmits  $a_{\text{in}}$ to $b_{\text{out}}$ if it is in state $R=0$.  
If the two-state model is driven by a poisson process with transition rates $\Gamma_{R\overline R}$, then the probabilities for the two states satisfy a rate equation 
\begin{align}
	\l[ \begin{array}{c}
		\dot P_0 \\
		\dot P_1
	\end{array} \r] &= \left[
	\begin{array}{cc}
		 -\Gamma _{01} &  \hphantom{-}\Gamma _{10} \\
		 \hphantom{-}\Gamma _{01} & -\Gamma _{10} \\
	\end{array}
	\right] \l[ \begin{array}{c}
		P_0 \\
		P_1
	\end{array} \r].
\end{align}
The steady-state probabilities are $p_0=\Gamma_{10}/\Gamma_{\text{tot}}$ and \mbox{$p_1=\Gamma_{01}/\Gamma_{\text{tot}}$}, where $\Gamma_{\text{tot}}=\Gamma_{01}+\Gamma_{10}$.  The return probabilities (i.e.\ the probability that if the system starts in state $R$, it will be found in the same sate after time $\tau$) is given by 
$P_{R,R}(\tau)=p_{R}-(1-p_{R})e^{-\Gamma_{\text{tot}} t}$.  Note that $P_{R,R}(0)=1$ and $\lim_{\tau\rightarrow\infty}P_{R,R}(\tau)=p_R$.  

This generic population rate model coincides with the rate model  \eqn{eqn:dgrate} when $P_0=P_D$, $P_1=P_G$,  $\Gamma_{01}=\delta^2\gamma$,  $\Gamma_{10}=2\delta^2\gamma$ and $\Gamma_{\text{tot}} = 3\delta^2\gamma$, leading to $p_1=p_G=1/3$, and $p_0=p_D=2/3$.

For a stationary process, and an incident coherent field, $\langle a_{\text{in}}(t)\rangle=\alpha_{\text{in}}$ it is straightforward to show that the reflected and transmitted  output field amplitudes and fluxes satisfy
\begin{align}
	\{\alpha_{\text{out}},\beta_{\text{out}}\}&=\{p_1\alpha_{\text{in}},p_0\alpha_{\text{in}} \},\nn\\
	\{\mathcal{A}_{\text{out}},\mathcal{B}_{\text{out}}\}&=\{p_1|\alpha_{\text{in}}|^2,p_0 |\alpha_{\text{in}}|^2 \},
\end{align}
where $\beta_{\text{out}}=\langle b_{\text{out}}(t)\rangle$,  $\mathcal{A}_{\text{out}} =	\langle a_{\text{out}}^\dagger(t) a_{\text{out}}(t) \rangle $ and $\mathcal{B}_{\text{out}} =	\langle b_{\text{out}}^\dagger(t) b_{\text{out}}(t) \rangle $.  
These reproduce the output field amplitudes and fluxes we have calculated for driving the two atom system from the coupled side (i.e.\ $\alpha$ driving) reported in \eqns{eqn:S} and (\ref{eqn:T}).

\begin{align}
	g_{\text{ref}}^{(1)}(\tau)& = \frac{\langle{a_{\text{out}}\hc(t+\tau) a_{\text{out}}(t)}\rangle }{{\langle{a_{\text{out}}\hc(t) a_{\text{out}}(t)}\rangle}},\nn\\
		&=\frac{\alpha_{\text{in}}^*P_{1,1}(\tau)\alpha_{\text{in}}p_1}{p_1 |\alpha_{\text{in}}|^2},\nn\\
		&=P_{1,1}(\tau),
\end{align}
and
\begin{align}
	g_{\text{trans}}^{(1)}(\tau)& = \frac{\langle{b_{\text{out}}\hc(t+\tau) b_{\text{out}}(t)}\rangle }{{\langle{b_{\text{out}}\hc(t) b_{\text{out}}(t)}\rangle}},\nn\\
		&=P_{0,0}(\tau).
\end{align}
These both satisfy $g_R^{(1)}(0)=1$, and decay exponentially to $g_R^{(1)}(0)=p_R$  as $\tau$ increases. Similarly
\begin{align}
	g_{\text{ref}}^{(2)}(\tau)& = \frac{\langle{a_{\text{out}}\hc(t)a_{\text{out}}\hc(t+\tau)a_{\text{out}}(t+\tau) a_{\text{out}}(t)}\rangle }{{\langle{a_{\text{out}}\hc(t) a_{\text{out}}(t)}\rangle^2}},\nn\\
		&=P_{1,1}(\tau)/p_1,
\end{align}
and
\begin{align}
	g_{\text{trans}}^{(2)}(\tau)& = \frac{\langle{b_{\text{out}}\hc(t)b_{\text{out}}\hc(t+\tau)b_{\text{out}}(t+\tau) b_{\text{out}}(t)}\rangle }{{\langle{b_{\text{out}}\hc(t) b_{\text{out}}(t)}\rangle^2}},\nn\\
		&=P_{0,0}(\tau)/p_0.
\end{align}
These both satisfy $g_R^{(2)}(0)=1/p_R$, and decay exponentially to unity as $\tau$ increases.

Identifying the rates in the toy model with the rate equation in Eq.~\eqref{eqn:dgrate}, we reproduce  the correlation functions for the $\alpha$-driven cascaded atom system. 

\end{document}